\documentclass[twocolumn, secnumaRabic, amssymb, nobibnotes, aps, showpacs,superscriptaddress]{revtex4-1}
\usepackage{graphicx,subfigure,amsmath}%
\usepackage{color}
\usepackage[bookmarks=false]{hyperref}
\hypersetup{colorlinks=true,citecolor=blue,linkcolor=blue,urlcolor=blue,pdfstartview=FitH,bookmarksopen=true}
\usepackage{amsmath,amstext}
\usepackage{graphicx}
\usepackage{booktabs}
\usepackage{multirow}
\usepackage{footmisc}
\definecolor{green}{rgb}{0.8,0.98,0.83}
\usepackage{float}
\usepackage{lipsum}

\usepackage{color}

\usepackage{ulem}

\newcommand{\stkout}[1]{\ifmmode\text{\sout{\ensuremath{#1}}}\else\sout{#1}\fi}

\begin{document}
\title{Complementary Collective Spin Descriptions of Superradiant Ramsey Spectroscopy 
}

\author{Ke-Xin Gao}
\address{School of Physics, Zhengzhou University, Zhengzhou 450052, China}

\author{Yuan Zhang}
\email{yzhuaudipc@zzu.edu.cn}
\address{School of Physics, Zhengzhou University, Zhengzhou 450052, China}
\address{Institute of Quantum Materials and Physics, Henan Academy of Sciences, Zhengzhou 450046, China}

\author{Shi-Lei Su}
\email{slsu@zzu.edu.cn}
\address{School of Physics, Zhengzhou University, Zhengzhou 450052, China}
\address{Institute of Quantum Materials and Physics, Henan Academy of Sciences, Zhengzhou 450046, China}

\author{Gang Chen}
\address{School of Physics, Zhengzhou University, Zhengzhou 450052, China}
\address{Institute of Quantum Materials and Physics, Henan Academy of Sciences, Zhengzhou 450046, China}

\author{Chongxin Shan}
\address{School of Physics, Zhengzhou University, Zhengzhou 450052, China}
\address{Institute of Quantum Materials and Physics, Henan Academy of Sciences, Zhengzhou 450046, China}

\author{Klaus M{\o}lmer}
\email{klaus.molmer@nbi.ku.dk}
\address{Niels Bohr Institute, University of Copenhagen, 2100 Copenhagen, Denmark}

\begin{abstract}
A recent experiment demonstrated delayed superradiance from strontium-88 atoms, which are coupled to a longitudinal mode of a cavity while being excited by laser pulses propagating along a transversal direction [Nat. Commun. 15, 1084 (2024)]. A coherent picture of the atomic ensemble dynamics in this experiment requires  complementary  representations of the external driving dynamics and the superradiant dynamics. To complement previous analyses, we introduce these representations by considering in-phase and out-of-phase superpositions of transverse collective spin components of two atomic sub-ensembles, and analyze the dynamics with the corresponding collective Dicke states and Bloch vectors. This approach  also explains Ramsey spectroscopy experiments with the ensemble,  and it may be  employed to explore other phenomena, such as  weak-to-strong coupling phase transitions and triggered superradiance.  
\end{abstract}

\maketitle

\section{Introduction}

In 1954, R. H. Dicke ~\citep{Dicke1954} presented a comprehensive analysis of  superradiance, i.e. cooperative spontaneous emission by many two-level atoms. For atoms confined well within  an optical wavelength and hence coupled uniformly to the radiation field, the so-called Dicke states can be used to describe the symmetric quantum states of the atoms, and to analyze the collective coupling of the atomic ensemble with the radiation field. Inspired by this seminal work, there were extensive theoretical and experimental studies in the 1980s on superradiance and related phenomena~\citep{AnatoliiVAndreev1980,Gross1982}. Since superradiance is generated by collective emission of the atomic ensemble, it was initially considered as a transient phenomenon.  However, in 2009,  Meiser et al. proposed~\citep{Meiser2009} that the collective decay can be compensated by incoherent excitation, and they predicted steady-state superradiance with ultra-narrow spectra in optical lattice clock systems. Because this radiation is robust against cavity fluctuations, it has strong application potential in quantum metrology and thus it has been studied extensively thereafter~\citep{Ludlow2015,PhysRevA.98.063837,Zhang2021,Bohnet2012,Norcia2016,Norcia2018}.

In many studies on superradiance, the coupling of  atoms to a single cavity mode has been treated as uniform giving rise to the essentially same physics as explored by Dicke. In 2006, Scully et al. introduced the so-called timed Dicke states~\citep{PhysRevLett.96.010501} to account for the spatial phase variation of the coupling, emphasizing the importance of phase-matching for laser excitation and collective emission. Recently, Hotter et al.~\cite{PhysRevResearch.5.013056} predicted that two atomic ensembles, which are driven coherently and in phase but are coupled with opposite phases to an optical cavity mode,  will display delayed superradiance. This was soon  demonstrated in experiments with strontium-88 atoms~\cite{Bohr2024}. These studies have opened a new direction in the research of collective effects, and prompt further studies on the influence of phase variation of the couplings in extended emitter systems. 

\begin{figure}[htbp]
\includegraphics[width=0.47\textwidth]{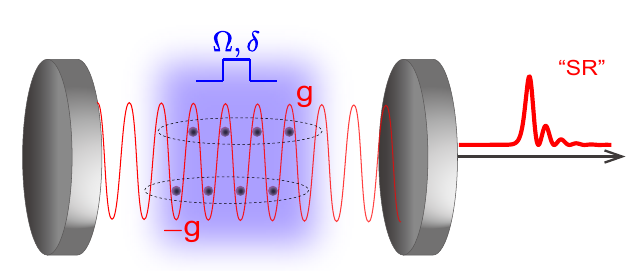}
\caption{\label{fig:system} Delayed superradiance from two atomic sub-ensembles, which are driven transversely by a laser pulse with a strength $\Omega \left ( t \right ) $ and a frequency detuning $\delta $, and couple with the optical cavity in a non-symmetric fashion with strengths $g, -g $. 
}
\end{figure}

In this article, we supplement the analyses in Ref.~\cite{PhysRevResearch.5.013056} by establishing two complementary pictures of the dynamics. One is based on the collective, uniform sum of raising and lowering operators that couple coherently to the laser field, the other based on the collective, staggered sum of raising and lowering operators that couple coherently to the cavity field. These sets of collective operators permit representations of the system dynamics in two different collective Dicke spin state spaces and in two different collective Bloch spheres.

In these two representations, the laser driving and the cavity coupling are particularly easy to identify, and we are able to establish a clear connection between the quantum state evolution and the final superradiant readout. Further applications of such dual representations may benefit analyses of the dependence of superradiance on the phase variation of the couplings, the preparation of atomic ensembles in specific superradiant or sub-radiant states, and the exploration of the weak-to-strong coupling phase transition~\citep{Zhang2022},  triggered superradiance~\citep{Kersten2023}, and metrology protocols such as heterodyne-based frequency measurements~\citep{Norcia2018,Zhang2022-1}.

The article is organized as follows. In  Sec.~\ref{sec:theory}, we present the theoretical model for the system, including the quantum master equation, its solution with the cumulant mean-field approach, and its interpretation by two complementary collective spin pictures. In Secs.~\ref{sec:delayed-superradiance} and ~\ref{sec:ramsey spectroscopy}, we use the collective spin pictures to analyze the dynamics of the delayed superradiance. In Sect.~\ref{sec:ramsey spectroscopy}, we simulate Ramsey spectroscopic measurements and discuss the possibility of using these measurements for frequency locking in atomic clocks. Finally, we summarize and provide an outlook for further investigations.

\section{\label{sec:theory} Theoretical Model}

In this section, we develop the quantum master equation to describe the system dynamics. We present its solution by a cumulant mean-field approach, and we introduce Dicke states and Bloch vectors in two complementary pictures.

\subsection{Quantum Master Equation\label{sec:qme}}

To account for  the inhomogeneous coupling of an atomic ensemble to a single cavity and the spatial profile of a classical driving field, we may divide the atomic ensemble into $N$ sub-ensembles, and assume that the $\alpha$-th sub-ensemble with $N_\alpha$ atoms couples with the cavity and the laser driving with the strength $g_\alpha,\Omega_\alpha$, respectively. The state of the system is then described by a reduced density operator, which follows a master equation incorporating all interactions between the system components and all dissipation. In a frame rotating with the driving laser frequency $\omega_d$ the quantum master equation has the form,
\begin{align}                          
& \frac{\partial}{\partial t}\hat{\rho} =-\frac{i}{\hbar}[\hat{H}_c + \sum_{\alpha=1}^N \sum_{k=1}^{N_{\alpha}}(\hat{H}_{\alpha,k}^{a} +  \hat{H}_{\alpha,k}^{a-c} +\hat{H}_{\alpha,k}^{a-d} ),\hat{\rho}]\nonumber \\
 & -\kappa\mathcal{D}[\hat{a}]\hat{\rho}-\sum_{\alpha=1}^N \sum_{k=1}^{N_{\alpha}}(\gamma_{\alpha}\mathcal{D}[\hat{\sigma}_{\alpha}^{12}]\hat{\rho}+2\chi_{\alpha}\mathcal{D}[\hat{\sigma}_{\alpha}^{22}]\hat{\rho}).\label{eq:meq}
\end{align}
In this equation, the Hamiltonian $\hat{H}_c = \hbar \delta_c \hat{a}^\dagger \hat{a}$ describes the optical cavity with photon creation $\hat{a}$ and annihilation operator $\hat{a}^\dagger$ and frequency detuning $\delta_c = \omega_c-\omega_d$. $\hat{H}_{\alpha,k}^a = \hbar \delta_\alpha  \hat{\sigma}_{\alpha,k}^{22}$ describes the excitation energy of the $k$-th atom of the $\alpha$-th sub-ensemble with projection operators $\hat{\sigma}_{\alpha,k}^{22}$ on the excited state and frequency detunings $\delta_\alpha = \omega_\alpha-\omega_d $. Here and in the following, we use the subscript $1,2$ in the atomic operators to indicate the ground and excited level, respectively. $\hat{H}_{\alpha,k}^{a-c} =\hbar  (g_{\alpha} \hat{a}^{\dagger}\hat{\sigma}_{\alpha,k}^{12}+ g_{\alpha}^* \hat{\sigma}_{\alpha,k}^{21}\hat{a})$ describes the coherent energy exchange between a single atom and the optical cavity, and $\hat{H}_{\alpha,k}^{a-d} =\hbar\Omega_{\alpha}(\hat{\sigma}_{\alpha,k}^{12}+\hat{\sigma}_{\alpha,k}^{21})$ describes the coherent driving by the laser field with transition operators $\hat{\sigma}_{\alpha,k}^{12},\hat{\sigma}_{\alpha,k}^{21}$. In Eq. (\ref{eq:meq}), the second line describes dissipation with the Lindblad superoperator $\mathcal{D}[\hat{o}]\hat{\rho}=\frac{1}{2}\left\{ \hat{o}^{\dagger}\hat{o},\hat{\rho}\right\} -\hat{o}\hat{\rho}\hat{o}^{\dagger}$ for any operator $\hat{o}$. The first term of this line describes the cavity photon loss with a rate $\kappa$, and the remaining terms describe the atomic spontaneous emission and dephasing with rates $\gamma_{\alpha},\chi_{\alpha}$. 

\subsection{Cumulant Mean-field Equations}\label{sec:cumulant}

To simulate the experimental system with thousands of atoms, a standard density matrix technique can not be explored due to the exponentially large  Hilbert space dimension. Instead, we adopt a cumulant mean-field approach~\citep{Plankensteiner2022}, and derive equations 
$\frac{\partial}{\partial t}\langle \hat{o}\rangle =\mathrm{tr}\left\{ \hat{o}\frac{\partial}{\partial t}\hat{\rho}\right\} $
for the mean values $\langle \hat{o}\rangle =\mathrm{tr}\left\{ \hat{o}\hat{\rho}\right\}$
of all relevant observables $\hat{o}$ from Eq. ~\eqref{eq:meq}. These equations include mean values of products of two operators $\langle \hat{o}\hat{p}\rangle $, which further depend on the mean values of products of three operators $\langle \hat{o}\hat{p}\hat{q}\rangle $
and so on. To truncate the hierarchy of equations, we employ the second-order cumulative expansion approximation $\langle \hat{o}\hat{p}\hat{q}\rangle \approx\langle \hat{o}\rangle \langle \hat{p}\hat{q}\rangle +\langle \hat{p}\rangle \langle \hat{o}\hat{q}\rangle +\langle \hat{q}\rangle \langle \hat{o}\hat{p}\rangle -2\langle \hat{o}\rangle \langle \hat{p}\rangle \langle \hat{q}\rangle $ to achieve a set of closed equations. If all the atoms in each sub-ensemble behave identically, the mean-values $\langle \hat{\sigma}_{\alpha,k}^{mn}\rangle ,\langle \hat{a}\hat{\sigma}_{\alpha,k}^{mn}\rangle ,\langle \hat{a}^{\dagger}\hat{\sigma}_{\alpha,k}^{mn}\rangle $
are the same for all atoms, denoted by $k$, and  $\langle \hat{\sigma}_{\alpha,k}^{mn}\hat{\sigma}_{\alpha',k'}^{mn}\rangle $ are same for all atom pairs, denoted by $k,k'$, inside one sub-ensemble ($\alpha=\alpha'$) and between two sub-ensembles ($\alpha\neq\alpha'$). This reduces the number of independent equations from the order of $(\sum_\alpha  N_{\alpha})^{2}$ to a few tens, irrespective of the number of atoms.

In our numerical calculations, we employ the QuantumCumulants.jl package~\citep{Plankensteiner2022} and we describe the corresponding codes in detail in the Appendix~\ref{sec:codes}. While 
the resulting full set of equations are summarized in  Appendix~\ref{sec:meanfields}, some examples are presented here. The intra-cavity photon number $\langle \hat{a}^\dagger \hat{a}\rangle$ satisfies the equation
\begin{equation}
\frac{d}{dt}\langle\hat{a}^\dagger\hat{a}\rangle = -\kappa\langle\hat{a}^\dagger\hat{a}\rangle + i\sum_{\alpha=1}^N N_\alpha g_\alpha\left(\langle\hat{a}\hat{\sigma}_{\alpha,1}^{21}\rangle -\langle\hat{a}^\dagger\hat{\sigma}_{\alpha,1}^{12}\rangle\right).
\end{equation}
The excited-state population $
\langle \hat{\sigma}_{\alpha,1}^{22}\rangle$ of the atoms satisfies the equation 
\begin{align}
&\frac d{dt}\langle\hat{\sigma}_{\alpha,1}^{22}\rangle =-\gamma_\alpha\langle\hat{\sigma}_{\alpha,1}^{22}\rangle+ig_\alpha\left(\langle\hat{a}^\dagger\hat{\sigma}_{\alpha,1}^{12}\rangle-\langle\hat{a}\hat{\sigma}_{\alpha,1}^{21}\rangle\right)\nonumber \\
&+i\Omega_\alpha\left(\langle\hat{\sigma}_{\alpha,1}^{12}\rangle-\langle\hat{\sigma}_{\alpha,1}^{21}\rangle\right),
\end{align}
and the atomic coherence $\langle\hat{\sigma}_{\alpha,1}^{12}\rangle$ satisfies the equation
\begin{align}
&\frac d{dt}\langle\hat{\sigma}_{\alpha,1}^{12}\rangle  =i\tilde{\delta}_a\langle\hat{\sigma}_{\alpha,1}^{12}\rangle+ig_\alpha\big(2\langle\hat{a}\hat{\sigma}_{\alpha,1}^{22}\rangle-\langle\hat{a}\rangle\big)\nonumber \\
&+i\Omega_\alpha\big(2\langle\hat{\sigma}_{\alpha,1}^{22}\rangle- 1\big). 
\end{align}
Here, we have introduced the complex frequency detunings $\tilde{\delta}_\alpha = \delta_\alpha+(\gamma_\alpha/2+\chi_\alpha)$. 
The atom-atom correlation in the same sub-ensemble $\langle \hat{\sigma}_{\alpha,1}^{12} \hat{\sigma}_{\alpha,2}^{21}\rangle$ satisfies the equation
\begin{align}
&\frac d{dt}\langle\hat{\sigma}_{\alpha,1}^{21}\hat{\sigma}_{\alpha,2}^{12}\rangle=- (\gamma_\alpha+2\chi_\alpha)\langle\hat{\sigma}_{\alpha,1}^{21}\hat{\sigma}_{\alpha,2}^{12}\rangle+ig_\alpha(\langle\hat{a}^\dagger\hat{\sigma}_{\alpha,1}^{12}\rangle \nonumber \\
&-\langle\hat{a}\hat{\sigma}_{\alpha,1}^{21}\rangle)+2ig_\alpha\big(\langle\hat{a}\hat{\sigma}_{\alpha,1}^{22}\hat{\sigma}_{\alpha,1}^{21}\rangle-\langle\hat{a}^\dagger\hat{\sigma}_{\alpha,2}^{12}\hat{\sigma}_{\alpha,1}^{22}\rangle\big)  \nonumber \\
& +i\Omega_\alpha\big(\langle\hat{\sigma}_{\alpha,1}^{12}\rangle-\langle\hat{\sigma}_{\alpha,1}^{21}\rangle\big) +2i\Omega_\alpha(\langle\hat{\sigma}_{\alpha,1}^{22}\hat{\sigma}_{\alpha,2}^{21}\rangle  -\langle\hat{\sigma}_{\alpha,1}^{22}\hat{\sigma}_{\alpha,2}^{12}\rangle), 
\end{align}
while the atom-atom correlation  between different sub-ensembles  $\langle\hat{\sigma}_{\alpha,1}^{21}\hat{\sigma}_{\alpha^{\prime},1}^{12} \rangle$ follows the equation
\begin{align}
&\frac d{dt}\langle\hat{\sigma}_{\alpha,1}^{21}\hat{\sigma}_{\alpha^{\prime},1}^{12} \rangle=i(\tilde{\delta}_{\alpha'} - \tilde{\delta}_\alpha)\langle\hat{\sigma}_{\alpha,1}^{21}\hat{\sigma}_{\alpha^{\prime},1}^{12}\rangle+ig_{\alpha}\langle\hat{a}^{\dagger}\hat{\sigma}_{\alpha^{\prime},1}^{12}\rangle\nonumber \\
&-ig_{\alpha^{\prime}}\langle\hat{a}\hat{\sigma}_{\alpha,1}^{21}\rangle  -2ig_\alpha\langle\hat{\sigma}_{\alpha,1}^{22}\hat{a}^\dagger\hat{\sigma}_{\alpha^{\prime},1}^{12}\rangle+2ig_{\alpha^{\prime}}\langle\hat{\sigma}_{\alpha^{\prime},1}^{22}\hat{a}\hat{\sigma}_{\alpha,1}^{21}\rangle\nonumber \\
& +i\Omega_\alpha\big(\langle\hat{\sigma}_{\alpha',1}^{12}\rangle-2\langle\hat{\sigma}_{\alpha,1}^{22}\hat{\sigma}_{\alpha',1}^{12}\rangle\big)+i\Omega_{\alpha'}(2\langle\hat{\sigma}_{\alpha,1}^{21}\hat{\sigma}_{\alpha',1}^{22}\rangle -\langle\hat{\sigma}_{\alpha,1}^{21}\rangle).
\end{align}

\subsection{Collective Dynamics in Two Complementary Pictures \label{sec:pictures}} 

\begin{figure}[htbp]
\includegraphics[width=0.49\textwidth]{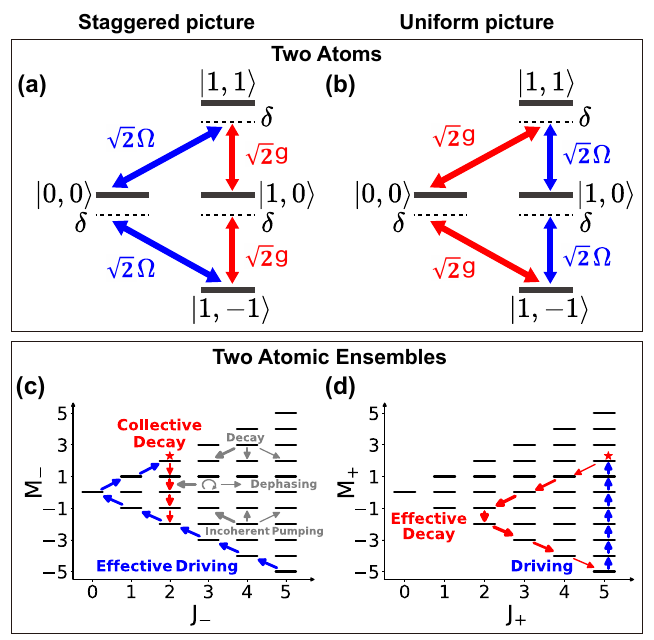}
\caption{\label{fig:couplings} Level diagram of collective Dicke states for systems with two (a,b) and ten atoms (c,d). In the staggered picture (a), the laser driving (blue arrows) couples the Dicke state $|J_- = 0,M_- = 0\rangle $, while the cavity mode (red arrows) couples $|J_- =1,M_- =0\rangle$. In the uniform picture (b), the laser driving (blue arrows) couples the Dicke state $|J_+ = 1,M_+ = 0\rangle $, while the cavity mode (red arrows) couples $|J_+ =0,M_+ =0\rangle$.
In the panels (c,d), the blue and red arrows indicate the processes among multi-atom  Dicke states due to the coupling with the laser and the cavity, resectively. In panel (c), the gray arrows show the flow of population due to atomic decay, dephasing and incoherent pumping~\citep{Zhang2018}. For the definition of the staggered and uniform picture, see the text.
}
\end{figure}

Although the master equation and the resulting mean-field equations can be readily applied with many atomic sub-ensembles, it is, for the purpose of this article, sufficient to consider two atomic sub-ensembles, i.e. $\alpha=1,2$. Due to the two sign values of the coupling amplitude to the cavity mode we refer to this coupling as staggered, while the coupling to the laser driving field is referred to as uniform. Correspondingly, we define first the collective spin operators for the two atomic sub-ensembles $\hat{J}_{\alpha,x(y)}=\frac{1(i)}{2}\sum_{k=1}^{N_{\alpha}}(\hat{\sigma}_{\alpha,k}^{12}\pm\hat{\sigma}_{\alpha,k}^{21})$, $\hat{J}_{\alpha,z}=\frac{1}{2}\sum_{k=1}^{N_{\alpha}}(2\hat{\sigma}_{\alpha,k}^{22}-\hat{1}_{\alpha,k})$ and, next, we consider the in-phase and out-phase superposition of the transverse components of the collective spin operator $\hat{J}^{\pm}_{i=x,y}= \hat{J}_{1,i} \pm \hat{J}_{2,i}$. With these operators, and the sum of the longitudinal components $\hat{J}^{\pm}_{i=z}= \hat{J}_{1,i}+\hat{J}_{2,i}$, we obtain two sets of collective spin operators and we can define the Dicke states of the total atomic ensemble as joint eigenstates, obeying $\sum_{i=x,yz}(\hat{J}_{i}^{\pm})^2 |J_\pm,M_\pm\rangle =J_\pm(J_\pm+1)|J_\pm,M_\pm\rangle $,
$\hat{J}^\pm_{z}|J_\pm,M_\pm\rangle =M_\pm|J_\pm,M_\pm\rangle $. The eigenvalues $J_\pm$ and $M_\pm$ are  integers or half-integers in the ranges, $J_\pm=0,...,(N_{1}+N_{2})/2$,$-J_\pm\le M_\pm\le J_\pm$. In addition, we introduce two collective Bloch vectors of the whole ensemble, $\mathbf{A}_\pm =\sum_{i=x,y,z} A_{\pm,i} \mathbf{e}_i=\sum_{i=x,y,z}  \langle \hat{J}_{i}^\pm \rangle \mathbf{e}_i$, where $\mathbf{e}_x,\mathbf{e}_y,\mathbf{e}_z$ are the unit vectors of the Cartesian coordinate system. As shown below, the association of the collective operators $\{\hat{J}^{-}_{i}\},\{\hat{J}^{+}_{i}\}$ with the staggered and uniform coupling make the quantities $J_-,M_-,\mathbf{A}_-$, and $J_+,M_+,\mathbf{A}_+$  very useful for the analysis of   the atomic ensemble dynamics.

At this stage, it is worthwhile to discuss the possible processes among the Dicke states in the two pictures. We start first with the simple case of just two atoms. By following the treatment in Appendix~\ref{sec:atomic_pairs}, we obtain the energy diagrams shown in Fig.~\ref{fig:couplings}(a,b). Figure ~\ref{fig:couplings}(a) shows that in the staggered picture, three triplet states including the Dicke state $|J_- = 1,M_-= 0\rangle$, are all coherently coupled by the cavity field, while the uniform laser beam excites the  state $|J_- = 0,M_- = 0\rangle$ of different symmetry. In contrast, Fig. ~\ref{fig:couplings}(b) shows that in the uniform picture, the three triplet states including the Dicke state $|J_+ = 1,M_+=0\rangle$, are all coherently coupled by the laser field, while the cavity field couples the state $|J_+ = 0,M_+ = 0\rangle$.

For more atoms (e.g. $10$ atoms), we obtain diagrams like the ones shown in Fig.~\ref{fig:couplings}(c,d). The Dicke states with different $M$ and given $J$ values form vertical ladders, and ladders for different $J$ are shifted horizontally, resulting in the triangular shapes. In analogy with the two atom case, in the staggered picture the laser excites the atomic ensemble along the lower and upper diagonal Dicke state boundaries, and the cavity field induces collective transitions among the Dicke states in the vertical direction. In contrast, 
in the uniform picture, the laser excites Dicke states in the vertical direction, and the cavity induces transitions along the diagonal boundaries of the Dicke states space. For reference, we note that in a single ensemble of two-level atoms, individual decay leads to jumps from $\left | J, M\right \rangle$ to the states $\left | J, M-1\right \rangle,\left | J\pm 1, M-1\right \rangle$, while incoherent pumping leads to upward jumps to the states with $M+1$, and individual dephasing leads to jumps into the same state and $\left | J\pm 1, M\right \rangle$~\citep{Meiser2009,Bohnet2012,Hotter2022,PhysRevA.98.063837,Zhang2018}.

To overcome the challenge of the large dimension of the multi-atom Hilbert space, which for millions of atoms remains prohibitively large, even when restricted to the symmetric Dicke states, we have recourse to the cumulant numerical method as explained in the previous section. However, to benefit from the intuitive dynamics in the  Dicke states space, we follow the procedure in Refs.~\citep{PhysRevA.98.063837,Zhang2021,Zhang2022} and define average Dicke state quantum  numbers $\bar{J}_\pm,\bar{M}_\pm$ through the equations $\bar{J}_\pm(\bar{J}_\pm+1)=\sum_{i=x,y,z}\langle (\hat{J}^\pm_{i})^{2}\rangle $ and $\bar{M}_\pm=\langle \hat{J}^\pm_{z}\rangle $. Thus, all the quantities $\bar{J}_\pm, \bar{M}_\pm, A_{\pm,i=x,y,z}$ introduced above can be computed with the mean values of single spin operators and products of single spin operators, which are available from the cumulant mean-field equations as described in the subsection~\ref{sec:cumulant}. The relevant formulas are summarized in the Appendix~\ref{sec:formula}. 

In our numerical simulations, unless otherwise stated, we shall use the parameters of the  experiment~\citep{Bohr2024}. The optical cavity has a frequency $\omega_c=2\pi\times435.1$ THz, and a photon damping rate $\kappa = 2\pi\times 0.78$ MHz, and couples with the atoms with the strength $g_{1} =-g_{2} = 2\pi\times 0.61 $ kHz. There are more than $N_\alpha = 10^7$ atoms in each sub-ensemble, and they all have the transition frequency $\omega_\alpha = \omega_c$, and the spontaneous emission rate $\gamma_\alpha = 2\pi \times 7.50$ kHz, and they are driven resonantly ($\delta_1=\delta_2=0$) by a laser with a strength parameterized by $\Omega_1 =\Omega_2 = 2\pi \times 4.16\times 10^5$ Hz. For simplicity, we ignore atomic dephasing ($\chi_\alpha =0$).  

\begin{figure}[htbp]
\includegraphics[width=0.5\textwidth]{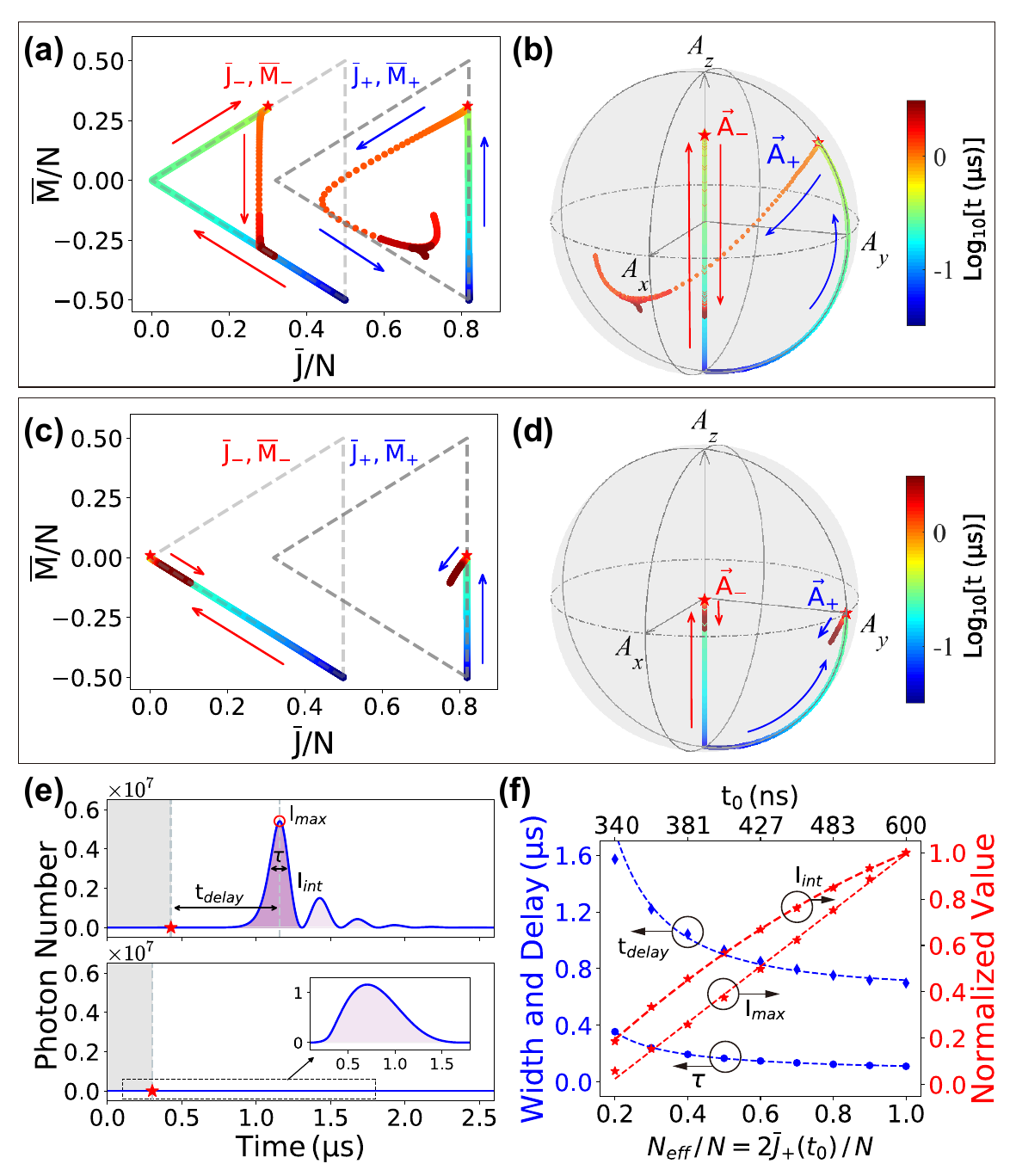}
\caption{\label{fig:pulses}  Atomic ensemble dynamics leading to delayed superradiance pulses.  Panel (a) shows the  dynamics of the atomic ensemble within the staggered and uniform Dicke state spaces, with quantum numbers $\bar{J}_-,\bar{M}_-$ (left) and $\bar{J}_+,\bar{M}_+$ (right), respectively. Here, the gray dashed lines show the boundaries of the Dicke states space, and the right Dicke states triangle is shifted horizontally for a better illustration. Panel (b) shows the corresponding dynamics of the collective Bloch vector in the staggered and uniformed picture, which are defined with $\mathbf{A}_-$ (evolving on the surface) and $\mathbf{A}_+$ (evolving along the z-axis), respectively.  Panel (c) and (d) show similar results for the system excited by a shorter driving pulse. In the above panels, the arrows indicate the direction of the time evolution. Panel (e) shows the mean intra-cavity photon number after a long (upper part) and short driving pulse (lower part), as marked by the light shaded areas. The dominant pulse is characterized by its maximum $I_{max}$, its integrated area $I_{int}$, its peak delay time $t_{delay}$, and width $\tau$ (as marked). Panel (f) shows  $I_{max},I_{int}$ (normalized to the largest value, right axis), $t_{delay},\tau $ (left axis) as functions of the effective number of atoms normalized by the total number $N_{eff}/N = 2\bar{J}_+(t_0)/N$} (lower axis) and  the driving pulse duration $t_0$ (upper axis), where $I_{max}, I_{int},t_{delay},\tau$ are well fitted with the expressions detailed in the main text.
\end{figure}

\section{Delayed Superradiance \label{sec:delayed-superradiance}}

To analyze the delayed superradiance, we consider the system dynamics in the presence of a laser driving pulse with a duration $0.427\,\mu s$, and show the time evolution in the staggered and uniform pictures in Fig.~\ref{fig:pulses}(a,b).  In the staggered  picture with the average Dicke numbers $\bar{J}_-,\bar{M}_-$,  the uniform laser excitation drives the atomic ensemble along the lower and then upper boundary of the Dicke states. Then, the excited atomic ensemble couples with the cavity field, and follows a vertical downward transition along the Dicke states with fixed $\bar{J}_-$, which is accompanied by the emission of a superradiant pulse as shown in the upper part of Fig.~\ref{fig:pulses}(e). 
If we, instead, consider the uniform picture with average Dicke states numbers $\bar{J}_+,\bar{M}_+$, the laser excitation makes the atomic ensemble climb the rightmost boundary of the Dicke states, and then the coupling with the cavity makes the ensemble explore the inner part of the Dicke state space in a seemingly incoherent manner. These dynamics agree with the pictures suggested in  Fig.~\ref{fig:couplings}(c) and (d).

We can illustrate the same dynamics with the collective Bloch vectors  [Fig.~\ref{fig:pulses}(b)]. In the staggered picture, the laser excitation makes the collective Bloch vector  $\mathbf{A}_-$ rise along the z axis from the south pole to a point below the north pole during the coherent driving, and subsequently the coupling with the cavity forces the Bloch vector to return along the axis to the south pole. In the uniform picture, however, during the coherent driving the collective Bloch vector $\mathbf{A}_+$ rotates around the x-axis from the south pole to a point above the equator, while it decays  downwards and leftwards through the Bloch sphere during the subsequent dynamics.

In Fig.~\ref{fig:pulses}(c,d) we show the system dynamics during and after a shorter driving pulse. The atomic ensemble moves downwards along the lower boundary of the $\bar{J}_-,\bar{M}_-$ Dicke states space in the staggered picture due to the individual atomic decay [Fig.~\ref{fig:system}(b)], and the delayed superradiance does not occur, as shown in the lower part of Fig.~\ref{fig:pulses} (e). 

The output intensity from the cavity is proportional to the intra-cavity mean photon number, as shown in Fig.~\ref{fig:pulses}(e). For the long driving pulse, we obtain a modulated output pulse  where the first peak is associated with the vertical Dicke states evolution shown in Fig.~\ref{fig:pulses}(a), and the subsequent modulation is caused by re-absorption and re-emission of cavity photons by the atomic ensemble, see Fig.~\ref{fig:rabi} in Appendix~\ref{sec:Extra}. Such fading pulse trains occur often in systems within the crossover coupling regime~\citep{Norcia2016}.  In contrast, with a short driving pulse, the intra-cavity photon number also rises and decreases, but with orders of magnitude lower values. To be more quantitative, we fit the dominant pulse component with a Gaussian function $f(t)=I_{max}{\rm exp}\{-4{\rm ln}2[t-(t_0+t_{delay})]^2/\tau^2\}$, and we characterize it with the maximum value $I_{max}$, the delay time $t_{delay}$ and the width $\tau$. Furthermore, we calculate also the pulse area $I_{int} = \int dt f(t) =  (\sqrt{\pi/{\rm ln}2}/2) I_{max} \tau$ with the product of the pulse maximum and duration. 

To obtain further insights into the formation of the delayed superradiant pulses, we define an effective number of atoms $N_{eff}=2\bar{J}_-(t_0)$ with the mean number $\bar{J}_-(t_0)$ of the Dicke states in the staggered picture at the end time  $t_0$  of the driving pulse.
In Fig.~\ref{fig:pulses}(f), we plot the characteristic parameters of the dominant pulses  as function of  the ratio $r=N_{eff}/N$. We find that $I_{max}$  can be well fitted by a linear dependence $I_{max}\approx I_{max,0}(1.21 r - 0.22)$ (with $I_{max,0}$ as the maximal value), and that $t_{delay}, \tau$ can be roughly fitted with the decreasing dependence, $t_{delay} \sim (0.08{\rm ln}/r+0.45)\times 10^{-6}, \tau \sim (0.05/r + 0.11)\times 10^{-6}$. As a result, the pulse area $I_{int} =  I_{max,0}\times1.51\times 10^{-2}\times(3.64 + 13.31 r- 1.10/r)$ is dominated by the scaling $\sim r$ for small $r<0.3$ and strongly affected by the scaling $\sim -1/r$ for large $r>0.3$, yielding the bending behavior of $I_{int}$ shown in Fig. ~\ref{fig:pulses}(e). 


The observed linear scaling of $I_{max}$ occurs since the radiation is not only affected  by the atomic dynamics in the Dicke states space, but also by the stimulated emission and absorption, which are particularly important for the system in the crossover or strong coupling regime~\citep{Gogyan2020}. As a comparison, we have also considered the system with fewer atoms in the weak coupling regime, and find that the intra-cavity photon number scales quadratically with  $N_{eff}$ (Fig.~\ref{fig:pulses-few-atoms}). In this case, all the scaling are consistent with the previous study on the superradiant pulses~\citep{Norcia2016}, and indicates the dominated influence of the collective decay.

\section{Ramsey Measurement with Delayed Superradiance \label{sec:ramsey measurement}}

\begin{figure}[htbp]
\includegraphics[width=0.49\textwidth]{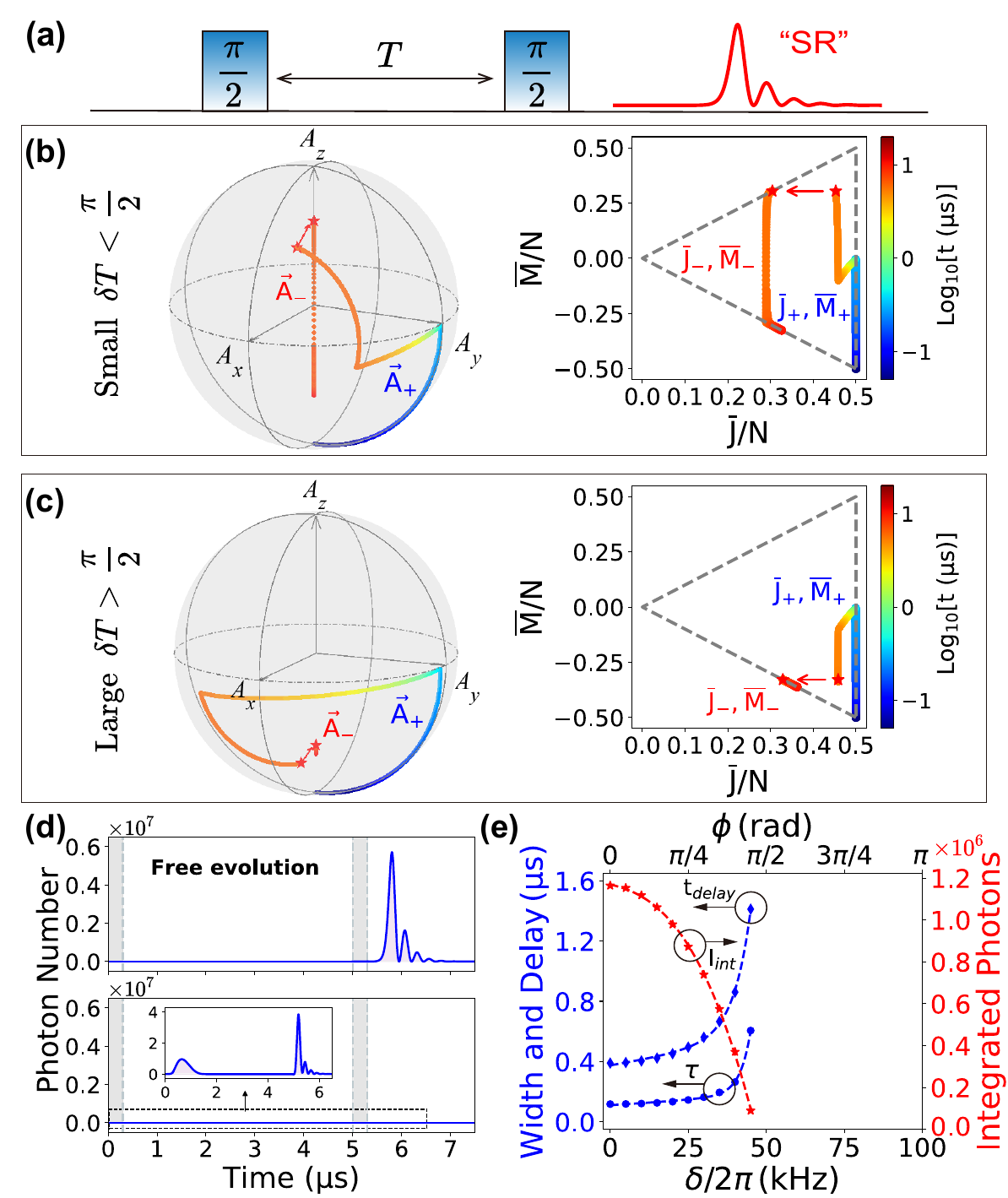}
\caption{\label{fig:Ramsey}
Ramsey measurement with two-pulse excitation and delayed superradiance. Panel (a) shows the Ramsey sequence: one $\pi/2$-pulse, a free-precession of duration $T$, another $\pi/2$-pulse, and readout through delayed superradiance. We assume that the atoms are resonant with the cavity, and are detuned from the driving laser by $\delta $. Panel (b) shows the presentation of the atomic ensemble dynamics with the Bloch vector and the Dicke states when the accumulated phase $\phi = \delta  T $ during the free-precision satisfies $\phi =\pi/4  $. Panel (c) shows the similar result but for   $\phi =3\pi/4$.  The dynamics of the collective Bloch vector is shown in a frame rotating with the driving field frequency. 
Panel (d) shows the intra-cavity photon number during the different phases of the dynamics for the cases with  $\phi = \pi/4$ (upper part)  and  $\phi = 3\pi/4 $ (lower part), respectively. Panel (e) shows the integrated pulse area $I_{int}$, the pulse delay $t_{delay}$ and the pulse width $\tau$ as functions of the frequency detuning $\delta$ (lower axis) and the accumulated phase $\phi$ in the range $[0,\pi]$ (upper axis). }
\end{figure}

After understanding the dynamics leading to the delayed superradiance, we study now the Ramsey measurement with such a signal (Fig.~\ref{fig:Ramsey}). The corresponding pulse sequence is shown in Fig.~\ref{fig:Ramsey}(a), which consists of one $\pi/2$ pulse, a free procession with duration $T$, another $\pi/2$ pulse, and finally a measurement of the superradiant signal through the photon-detector. Inspired by the analysis in the previous section, we utilize the quantities ${\bf A}_-, \bar{J}_-, \bar{M}_-$ and ${\bf A}_+, \bar{J}_+, \bar{M}_+$ to visualize the dynamics due to the laser excitation and the superradiant pulses, respectively, and introduce a mapping between them at the end of the free procession. Here, we assume that the atoms are resonant with the cavity, but are detuned from the driving field by $\delta $. To illustrate the dynamics with the collective Bloch vector, we consider the frame rotating with the laser frequency. Thus, during the free-precision, the Bloch vector will accumulate a phase $\phi = \delta  T$. In the following, we distinguish two situations with $\phi =\pi/4 \in [0,\pi/2]$ and $\phi =3\pi/4 \in [\pi/2,3\pi/2]$ in Fig.~\ref{fig:Ramsey}(b) and (c), respectively. 

In the  case with  $\phi =\pi/4$, following the protocol in Fig.~\ref{fig:Ramsey}(a), the collective Bloch vector $\mathbf{A}_+$ behaves as follows [left of Fig.~\ref{fig:Ramsey}(b)]. It rotates first around the x-axis by the angle $\pi/2$,  and rotates then clock-wise around the z-axis with slightly vertical decline. Then, the Bloch vector rotates around the x-axis by the angle $\pi/2$ again to one point in the x-z plane. After mapped to the vector $\mathbf{A}_-$ along the z-axis with the same z-coordinate, the Bloch vector decays vertically to the point over the the south pole. Correspondingly, in the Dicke states space  [right of Fig.~\ref{fig:Ramsey}(b)], the atomic ensemble climbs first along the rightmost boundary of the Dicke states space, then it declines towards smaller values of $\overline{J}_-,\overline{M}_-$, and then it climbs upwards along a line parallel to the rightmost boundary. Finally it maps to the Dicke states with $\overline{J}_+,\overline{M}_+$  along the upper boundary,  and decays vertically to the lower boundary of the Dicke triangle. 

In the latter case with $\phi =3\pi/4$, the system dynamics is similar except that the collective Bloch vector rotates around the z-axis for a much larger angle during the free-procession, and then it is rotated to the part of the x-z plane below the equator, and finally it is projected to the point along the z-axis below the equator [left part of Fig.~\ref{fig:Ramsey}(c)]. Accordingly, the atomic ensemble moves downwards along a line parallel to the rightmost  boundary during the second $\pi/2$ pulse, and it is mapped to the Dicke states on the lower boundary [right part of Fig.~\ref{fig:Ramsey}(c)]. 

According to the discussion of the dynamics displayed in Fig.~\ref{fig:pulses}, we expect that the former case would lead to a superradiant pulse and the latter would  not, as confirmed in Fig.~\ref{fig:Ramsey}(d). When the frequency detuning varies continuously, the Ramsey procedure maps the accumulated phase during the free precession to the emitted pulse area, pulse width and delay time, which is illustrated in Fig.~\ref{fig:Ramsey}(e) by their dependence  on the detuning $\delta$ (lower axis) and accumulated phase $\phi$ (upper axis). 

In short, the use of the collective Bloch vector and the Dicke states in two complementary pictures  are clearly useful to portray the whole process and explain the features observed in Fig. 3(a) in the experimental article~\citep{Bohr2024}.

\begin{figure}[htbp]
\includegraphics[scale=0.42]{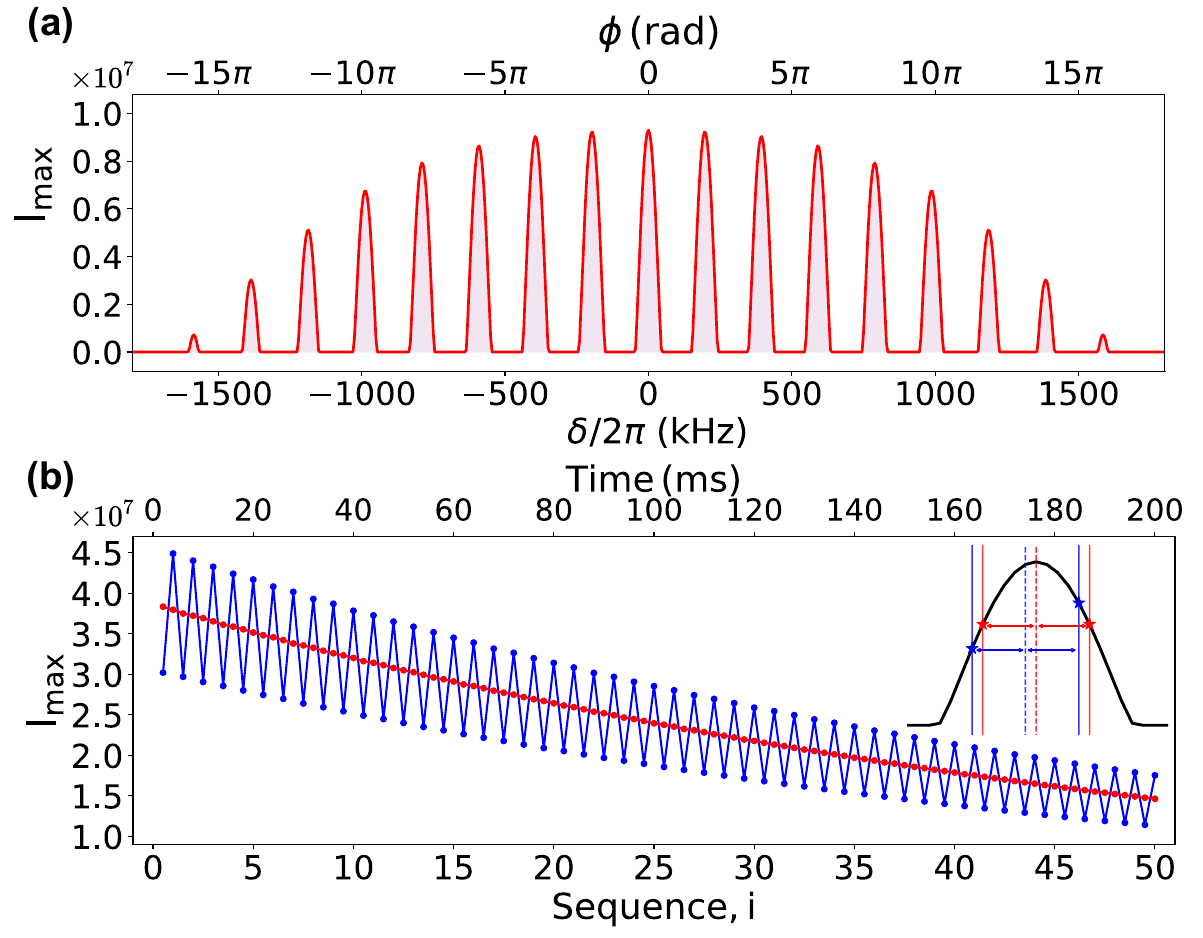}
\par
\caption{\label{fig:spectroscopy} Ramsey spectroscopy for longer accumulated phase. Panel (a) shows the maximal photon number $I_{max}$ of the superradiant signal as a function of the frequency detunning $\delta$ (lower axis) or the accumulated phase $\phi=\delta  T$ (upper axis) during the free-procession. Panel (b) mimics the key ingredient of frequency locking mechanism in the optical atomic clock, where  $I_{max}$ for two frequency detunings, which are either symmetric or non-symmetrical to the center Ramsey line (inset), is shown as function of the simulation sequence (lower axis) and the time (upper axis).  Here, the decay of the signal is caused by the atomic loss. For more details, see the text. 
} 
\end{figure}

\section{Ramsey Spectroscopy with Large Accumulated Phases \label{sec:ramsey spectroscopy}}

So far, we analyzed the system dynamics and the Ramsey measurement signal  with the accumulated phases $\phi = \delta  T $  within the range $[0,\pi]$. In Fig.~\ref{fig:spectroscopy}, we show results for phases in a broader range of values. By following the dynamics in Fig.~\ref{fig:Ramsey}(a), we expect that in the ideal case the collective Bloch vector will rotate clockwise and explore all directions around the z axis during the free precession. The second $\pi/2$ pulse will rotate the collective Bloch vector to directions above or below the equator, which eventually leads to the observable superradiant signal or no signal as shown in Fig.~\ref{fig:spectroscopy}(a). 

A detailed analysis of the freqeuncy detuning dependence in Fig.~\ref{fig:spectroscopy}(a) indicates that the peaks and zeros are not characterized by phases that are exact multiples of $\pi$. In fact, the intensity peaks (zeros)  amount to smaller (larger) azimuthal angle intervals of the collective Bloch vector as the frequency detuning $\delta$ is increased.  This occurs because the driving pulses are off-resonant and hence lead to imperfect $\pi/2$ pulses. The collective Bloch vector is thus rotated to directions below the equator plane by the first  pulse, and the degree of excitation after the second pulse has a more complex dependence on the accumulated phase between the pulses, see Fig.~\ref{fig:Pulse-large} in the Appendix~\ref{sec:Extra}.

The article~\citep{Bohr2024} also demonstrated the potential application of delayed superradiance for frequency locking. Inspired by this experiment and with our understanding of the Ramsey spectroscopy, we simulate now the locking of an optical signal to the central Ramsey frequency peak. To this end, we assume pairs of Ramsey processes with a driving laser of frequencies  $\omega_r+\delta$ and $\omega_r-\delta$. We repeat the simulations to obtain the signals as function of simulation cycles, and we compare the results for the resonant ($\omega_r=\omega_a$) and off-resonant case ($\omega_r\neq\omega_a$)  [Fig.~\ref{fig:spectroscopy}(c)]. To model the experiment, we assume that  each cycle lasts over $4$ ms ($2$ ms for one signal), which is mainly used for the atomic cooling, and the  number of atoms $N=N_0 e^{-\gamma_{loss}t}$ decreases exponentially with an initial value $N_0 = 4.47\times 10^7$ and a loss rate $\gamma_{loss} = 0.345$ Hz.  We find that the Ramsey signal reduces exponentially but smoothly with time for the resonant case,  while it shows a zig-zag structure for the off-resonant case.  These results agree   quantitatively with Fig. 4 of the experiment ~\citep{Bohr2024}.  By using the two signals in each cycle, one can generate an error signal and lock the reference laser frequency $\omega_r$.  With incorporation of all physical noise processes, it might be also possible to predict the frequency precision of the atomic clocks and guide further experimental exploration.

\section{Conclusions} 

In summary, we have developed a unified picture to reveal the physics involved in a recent experiment, which demonstrated the superradiance-based Ramsey spectroscopy with two atomic sub-ensembles, which are coupled with a driving field and a cavity in a in-phase and out-phase manner. To this end, we established two complementary pictures by considering  the in-phase and out-phase superposition of the transverse collective spin operators of the atomic ensembles, utilizing the two pictures to illustrate the laser driving and the cavity-mediated decay with the Dicke states and the collective Bloch vector, separately, and mapping the two dynamics at the end of the laser driving. Within this unified picture, we established a clear connection between the quantum states evolution and the detected superradiant pulses, and provided deep insights into the superradiant Ramsey spectroscopy.  In future,  the unified picture might be used to explore in more details the dependence of the superradiance on the relative phase of the couplings and the number of involved atomic sub-ensembles. The revealed mechanism can be also used to prepare the atomic ensemble to the specific superradiant and sub-radiant states, and to study  weak-to-strong coupling phase transition~\citep{Zhang2022},  triggered superradiance~\citep{Kersten2023}, quantum measurement backaction~\citep{Zhang2022-1} .

\begin{acknowledgments}
KeXin Gao carried out the numerical calculations under the supervision of Yuan Zhang, who developed the theory and the numerical programs. They contribute equally to the work. All authors contributed to the analyses and the writing of the manuscript. This work is supported the National Key R\&D Program of China under grant 2024YFE0105200, 
by  the National Natural Science Foundation of China under the grants 12422413 and 62475242,  Beijing National Laboratory for CondensedMatter Physics under the grant
 2023BNLCMPKF017, and the Cross-disciplinary Innovative Research Group Project of Henan Province No. 232300421004, as well as by the Carlsberg Foundation through the ``Semper Ardens'' Research Project QCooL. 
\end{acknowledgments}

\bibliography{references}

\newpage
\newpage
\appendix
\renewcommand\thefigure{A\arabic{figure}}
\renewcommand\thetable{A\arabic{table}}
\setcounter{figure}{0}

\section{\label{sec:codes}Julia Codes}

\begin{figure}[htbp]
\begin{centering}
\includegraphics[width=0.5\textwidth]{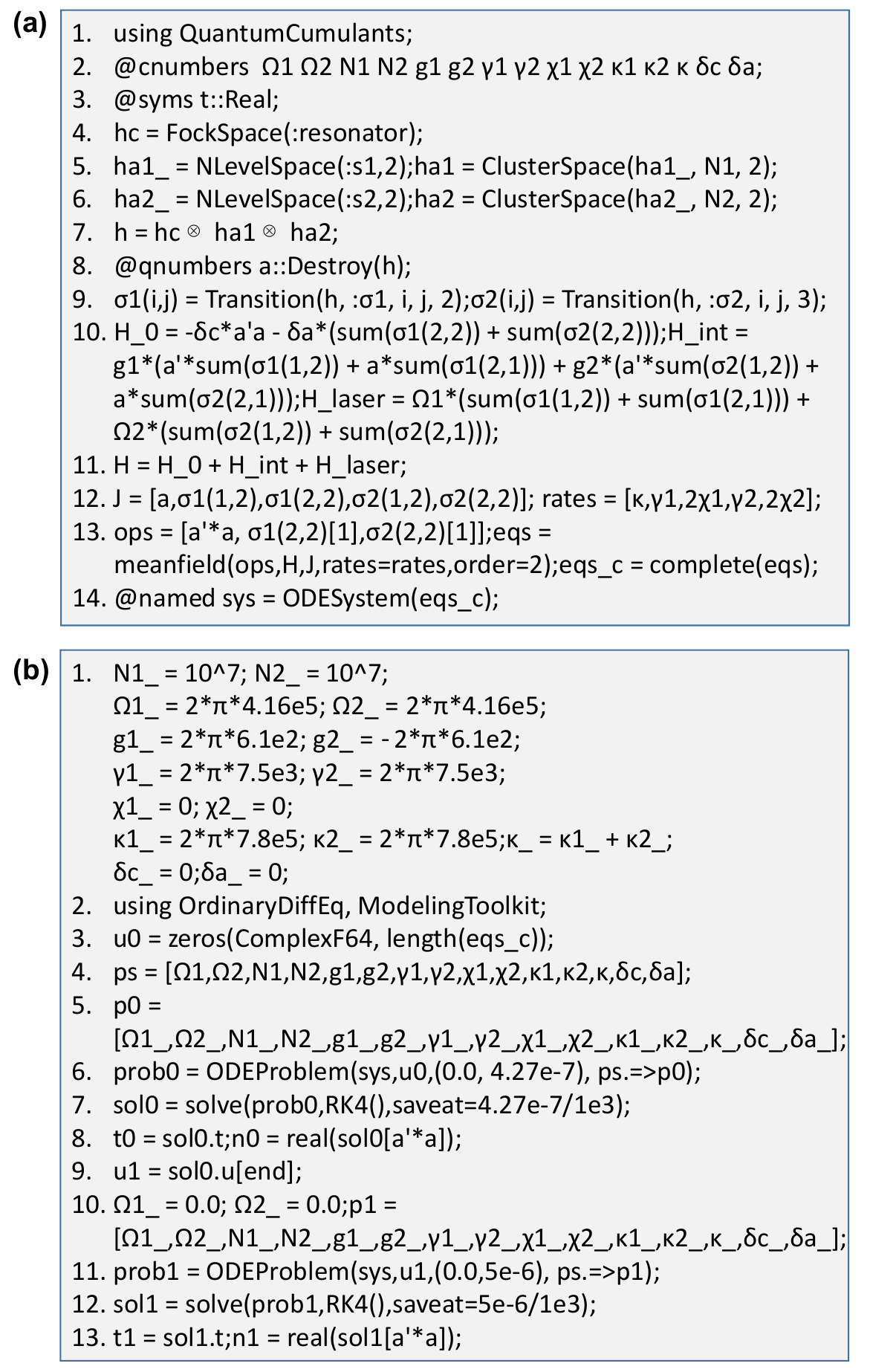} 
\par\end{centering}
\caption{\label{fig:codes}Julia codes to derive the mean-field equations (a) from the quantum master equation, and to solve the equations numerically (b). The details of the codes are explained in the text. }
\end{figure}

\begin{figure}[htbp]
\begin{centering}
\includegraphics[width=0.48\textwidth]{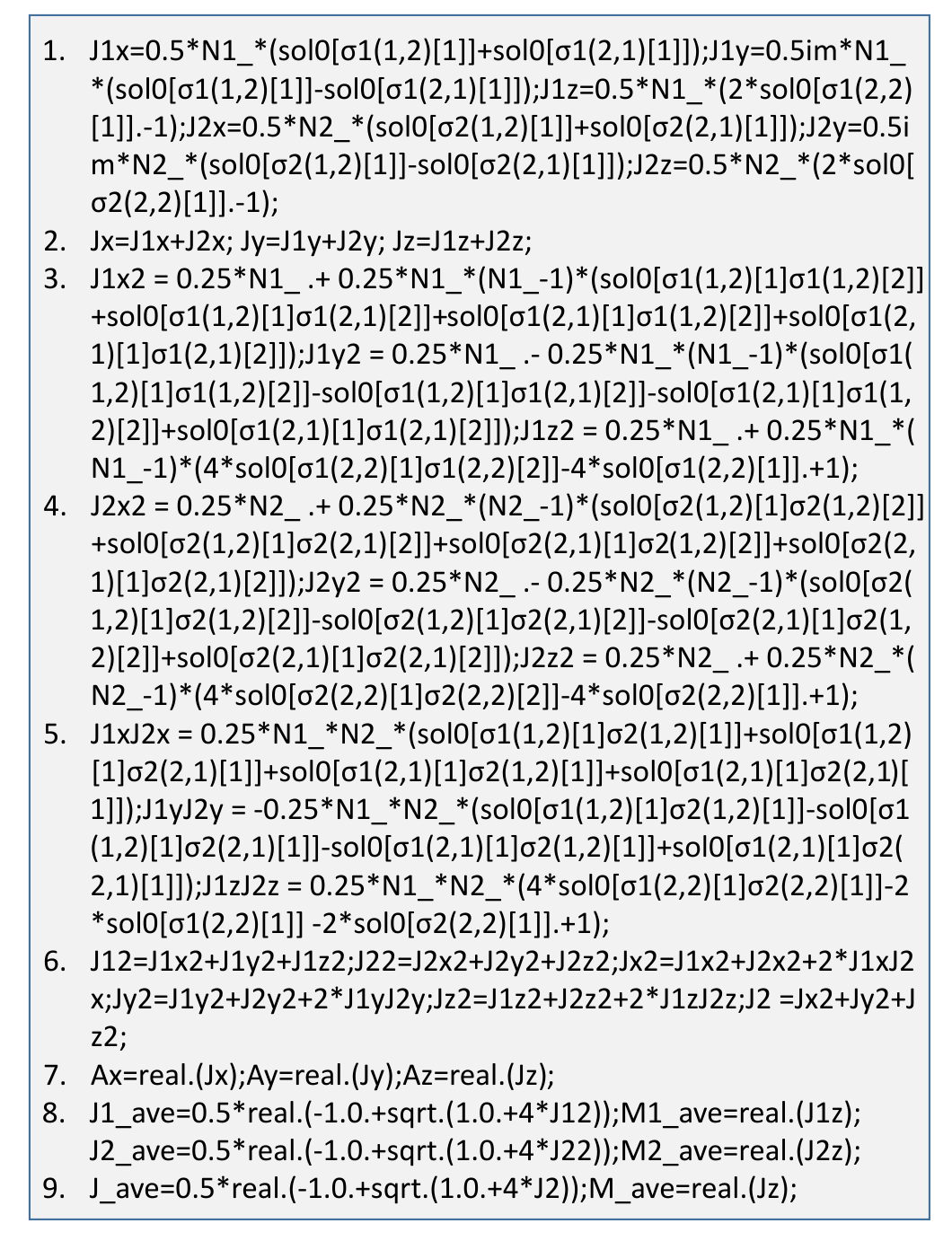} 
\par\end{centering}
\caption{\label{fig:codes2} Juila code for calculating the Dicke state quantum number and collective Bloch vector when the collective spin angular momenta of two sub-ensembles are combined in-phase.The details of the codes are explained in the text.}
\end{figure}

In this Appendix, we present the Julia codes to derive the mean-field equations [Fig.~\ref{fig:codes}(a)], and solve these equations numerically [Fig.~\ref{fig:codes}(b)], as well as the codes to compute the average of the Dicke states quantum numbers and the collective Bloch vector [Fig.~\ref{fig:codes2}]. 

In  Fig. \ref{fig:codes}(a), the 1st line imports the QuantumCumulants.jl packages.  The 2nd and 3rd line define the complex number and the symbol for the time. The 4th to 7th lines define the Hilbert space for the sub-system, e.g. the optical cavity as a quantized harmonic oscillator, the atoms as two-level systems, the atomic ensemble, and then define the product Hilbert space for the total system. The 8th and 9th line define the photon annihilation operator, and the transition and projection operator of the atoms. The 10th and 11th line define the Hamiltonian in the frame rotating with the driving laser frequency. The 12th line defines the list of operators and rates, which are used to specify the Lindblad terms. The 13th line defines the list of operators, derives the equations for the mean-values of these operators, analyzes the unknown quantities and derives the equations for them to form a closed set of equations.The 14th line defines the ordinary differential equation (ODE) problem. 

In Fig.~\ref{fig:codes}(b), the 1st line specifies the values of the parameters. The 2nd line imports the OrdinaryDiffEq.jl and ModelingToolkit.jl packages. The 3rd line defines the initial value of the mean-field quantities. The 4th and 5th line define the list of parameters and their values. The 6th and 7th line define the ODE problem, and solve this problem numerically with Runge-Kutta method. The 8th line calculates the simulation time and the mean intra-cavity photon number for system in the presence of the laser driving pulse. The 9th to 13th line are similar to the lines 3 to 8, and they calculate the evolution of the mean intra-cavity photon number after the laser driving pulse. 

In Fig.~\ref{fig:codes2}, the 1st line calculates the expectation values of the collective operators for two atomic sub-ensemble, while the 2nd line computes the value for the total ensemble. The 3rd and 4th line evaluate the expectation values of the squared collective operators for the first and second atomic ensemble, and the 5th line determines the expectation values of the product of the collective operators from the two sub-ensembles. The 6th line computes the expectation values of the squared collective operator of the total ensemble. The 7th line computes the collective Bloch vector  ${\bf A}_+$ of the total ensemble. The last two lines compute the average of Dicke state quantum numbers $\bar{J}_+,\bar{M}_+$ for the sub-ensemble and the total ensemble $\bar{J}_+,\bar{M}_+$. Here, we have presented the codes to compute ${\bf A}_+$ and $\bar{J}_+,\bar{M}_+$. However, these codes can be easily modified to compute  ${\bf A}_-$ and $\bar{J}_-,\bar{M}_-$.

\section{Level Diagram of an Atomic Pair \label{sec:atomic_pairs}}

In the subsection~\ref{sec:pictures} of the main text, we have introduced an energy diagram for the simplest system with two two-level atoms, and then analyzed the couplings with the laser and cavity. In this Appendix, we explain how this diagram is constructed. With the transition operators $\hat{\sigma}_{\pm}^{\dagger}=\frac{1}{\sqrt{2}}(\hat{\sigma}_{1}^{21}\pm\hat{\sigma}_{2}^{21})$, we can reformulate the  Hamiltonians involving atomic operators in Eq.\eqref{eq:meq}  as $\hat{H}^a = \xi_+ \sum_{\beta=\pm}\hat{\sigma}^{\dagger}_\beta\hat{\sigma}^{-}_\beta + \xi_- \sum_{\beta=\pm}\hat{\sigma}^{\dagger}_\beta \hat{\sigma}^{-}_{\bar{\beta}}$, $\hat{H}^{a-d}=\hbar\sum_{\beta=\pm}\Omega_{\beta}(\hat{\sigma}_{\beta}+\hat{\sigma}_{\beta}^{\dagger})$, $\hat{H}^{a-c}=\hbar\sum_{\beta=\pm}g_{\beta}(\hat{a}^{\dagger}\hat{\sigma}_{\beta}+ \hat{\sigma}_{\beta}^{\dagger}\hat{a})$ with the coefficients $\xi_\pm = \frac{1}{2}(\delta_1 \pm \delta_2)$,  $\Omega_{\pm}=\frac{1}{\sqrt{2}}\left(\Omega_{1} \pm\Omega_{2}\right)$, $g_{\pm}=\frac{1}{\sqrt{2}}(g_{1} \pm g_{2})$, respectively. Note that $\bar{\beta}=\mp$ for $\beta=\pm$. 

For the setting of the parameters $\delta_1=\delta_2=\delta, \Omega_{1}=\Omega_{2}=\Omega$ and $g_{1}=-g_{2}=g$ as explored in Ref.~\cite{PhysRevResearch.5.013056} and  ~\cite{Bohr2024}, we obtain $\xi_+=\delta,\xi_-=0$, $\Omega_+=\sqrt{2}\Omega,\Omega_-=0$ and $g_+=0,g_-=\sqrt{2}g$. Then, the above Hamiltonians can be simplified as $\hat{H}^a = \hbar \delta \sum_{\beta=\pm}\hat{\sigma}^{\dagger}_\beta\hat{\sigma}^{-}_\beta$, $\hat{H}^{a-d}=\hbar \sqrt{2}\Omega(\hat{\sigma}_{+}+\hat{\sigma}_{+}^{\dagger})$, $\hat{H}^{a-c}=\hbar \sqrt{2}g(\hat{a}^{\dagger}\hat{\sigma}_{-}+ \hat{\sigma}_{-}^{\dagger}\hat{a})$. By acting with the first Hamiltonian on the in-phase and out-phase singly excited states $|e_+\rangle =(|e_1,g_2\rangle + |g_1,e_2\rangle)/\sqrt{2},|e_- \rangle=(|e_1,g_2\rangle - |g_1,e_2\rangle)/\sqrt{2}$, of the atomic pair, we obtain $\hat{H}^a |e_+ \rangle = \hbar \delta |e_+ \rangle, \hat{H}^a |e_- \rangle = \hbar \delta |e_- \rangle $, and thus the transitions between these states and the ground state $|g_1 \rangle|g_2\rangle$ and the excited state are detuned from the laser driving by $\delta$. By acting with the second Hamiltonian on the ground state and the singly excited state $|e_+\rangle$ of the atomic pair, we obtain $\hat{H}^{a-d}|g_1\rangle|g_2\rangle = \hbar \sqrt{2}\Omega |e_+\rangle$, $\hat{H}^{a-d}|e_+\rangle = \hbar \sqrt{2}\Omega |e_1\rangle|e_2\rangle$, and find that the laser driving couples with the transitions related to the state $|e_+\rangle$ with a strength $\sqrt{2}\Omega$. By acting with the third Hamiltonian on the ground state and the singly excited state $|e_-\rangle$ of the atomic pair, we obtain $\hat{H}^{a-c}|g_1\rangle|g_2\rangle=\hbar \sqrt{2} g \hat{a} |e_-\rangle$, $\hat{H}^{a-c}|e_-\rangle=\hbar \sqrt{2} g \hat{a} |e_1\rangle |e_2\rangle$, and find that the cavity couples with the transitions related to the state $|e_-\rangle$ with a strength $\sqrt{2}g$.

The energy diagram in the staggered picture, as shown in Fig.~\ref{fig:couplings} (a), can be established by associating the general ground and excited state $| g_1,g_2\rangle$, $| e_1,e_2\rangle$ with the Dicke states $|J_-=1,M_-=-1\rangle$, $|J_-=1,M_-=1\rangle$, and the in-phase and out-phase entangled states $|e_+\rangle\rangle,|e_-\rangle\rangle$ with the Dicke states  $|J_-=0,M_-=0\rangle$ $|J_-=1,M_-=0\rangle$. Correspondingly, the energy diagram in the uniform picture, as shown in Fig.~\ref{fig:couplings} (b), can be constructed with a similar treatment except that the two states  $|e_+\rangle\rangle,|e_-\rangle\rangle$ are now assigned to the Dicke states  $|J_+=1,M_+=0\rangle$ $|J_+=0,M_+=0\rangle$.


\section{Calculation of Quantities in Complementary Pictures\label{sec:formula}}

In the subsection~\ref{sec:pictures} of the main text, we have introduced two complementary pictures to interpret the dynamics of the whole atomic ensemble. In these pictures, we have introduced the average of Dicke states quantum numbers $\bar{J}_\pm, \bar{M}_\pm = \langle \hat{J}^\pm_{z}\rangle$ and the collective Bloch vectors $\mathbf{A}_\pm =\sum_{i=x,y,z}A_{\pm,i}\mathbf{e}_i = \sum_{i=x,y,z}\langle \hat{J}_{i}^\pm \rangle\mathbf{e}_i$, where $\bar{J}_\pm$ are solutions of the equations $\bar{J}_\pm(\bar{J}_\pm+1)=\sum_{i=x,y,z}\langle (\hat{J}^\pm_{i})^{2}\rangle$. Obviously, these quantities are determined by the mean values of the collective operators $\hat{J}^\pm_i$ and the squared operators $(\hat{J}^\pm_i)^2$.

Assuming that the atoms are identical in sub-ensembles, we can calculate the required mean fields with individual atoms and atomic pairs with the following expressions:
\begin{align}
&\langle \hat{J}_{\alpha,x} \rangle = \frac{1}{2} N_{\alpha} (\langle \hat{\sigma}_{\alpha,1}^{12} \rangle + \langle \hat{\sigma}_{\alpha,1}^{21} \rangle), 
\end{align}
\begin{align}
&\langle \hat{J}_{\alpha,y} \rangle = \frac{i}{2} N_{\alpha} (\langle \hat{\sigma}_{\alpha,1}^{12} \rangle - \langle \hat{\sigma}_{\alpha,1}^{21} \rangle),
\end{align}
\begin{align}
&\langle \hat{J}_{\alpha,z} \rangle = \frac{1}{2} N_{\alpha} (2 \langle \hat{\sigma}_{\alpha,1}^{22} \rangle - 1),
\end{align}
\begin{align}
&\langle \hat{J}_{\alpha,x(y)}^2 \rangle = \frac{1}{4} N_{\alpha} \pm \frac{1}{4} N_{\alpha}(N_{\alpha} - 1)(\langle \hat{\sigma}_{\alpha,1}^{12} \hat{\sigma}_{\alpha,2}^{12} \rangle \pm \langle \hat{\sigma}_{\alpha,1}^{12} \hat{\sigma}_{\alpha,2}^{21} \rangle \nonumber \\
&\pm \langle \hat{\sigma}_{\alpha,1}^{21} \hat{\sigma}_{\alpha,2}^{12} \rangle + \langle \hat{\sigma}_{\alpha,1}^{21} \hat{\sigma}_{\alpha,2}^{21} \rangle), 
\end{align}
\begin{align}
&\langle \hat{J}_{\alpha,x(y)} \hat{J}_{\alpha',x(y)} \rangle = \pm \frac{1}{4} N_{\alpha} N_{\alpha'} (\langle \hat{\sigma}_{\alpha,1}^{12} \hat{\sigma}_{\alpha',1}^{12} \rangle \pm \langle \hat{\sigma}_{\alpha,1}^{12} \hat{\sigma}_{\alpha',1}^{21} \rangle \nonumber \\
&\pm \langle \hat{\sigma}_{\alpha,1}^{21} \hat{\sigma}_{\alpha',1}^{12} \rangle + \langle \hat{\sigma}_{\alpha,1}^{21} \hat{\sigma}_{\alpha',1}^{21} \rangle), 
\end{align}
\begin{align}
&\langle \hat{J}_{\alpha,z}^2 \rangle = \frac{1}{4} N_{\alpha} + \frac{1}{4} N_{\alpha}(N_{\alpha} - 1)(4 \langle \hat{\sigma}_{\alpha,1}^{22} \hat{\sigma}_{\alpha,2}^{22} \rangle - 4 \langle \hat{\sigma}_{\alpha,1}^{22} \rangle + 1), 
\end{align}
\begin{align}
&\langle \hat{J}_{\alpha,z} \hat{J}_{\alpha',z} \rangle = \frac{1}{4} N_{\alpha} N_{\alpha'} (4 \langle \hat{\sigma}_{\alpha,1}^{22} \hat{\sigma}_{\alpha',1}^{22} \rangle - 2 \langle \hat{\sigma}_{\alpha,1}^{22} \rangle - 2 \langle \hat{\sigma}_{\alpha',1}^{22} \rangle + 1).
\end{align}

\section{Second-order Mean-field Equations ~\label{sec:meanfields}}

In this Appendix, we present the derived second-order mean-field equations. We encounter three first-order mean-fields, i.e. the intra-cavity field amplitude $\langle\hat{a}\rangle$, the atomic coherence $\langle\hat{\sigma}_{\alpha,1}^{12}\rangle$, the atomic population $\langle\hat{\sigma}_{\alpha,1}^{22}\rangle$. These mean-fields satisfy the following equations  
\begin{align}
&\frac{d}{dt}\langle\hat{a}\rangle = i\tilde{\delta}_c\langle\hat{a}\rangle - i\sum_{\alpha=1,2} N_\alpha g_\alpha\langle\hat{\sigma}_{\alpha,1}^{12}\rangle, \label{eq:B9}
\end{align}
\begin{align}
&\frac d{dt}\langle\hat{\sigma}_{\alpha,1}^{12}\rangle  =i\tilde{\delta}_a\langle\hat{\sigma}_{\alpha,1}^{12}\rangle+ig_\alpha\big(2\langle\hat{a}\hat{\sigma}_{\alpha,1}^{22}\rangle-\langle\hat{a}\rangle\big)\nonumber \\
&+i\Omega_\alpha\big(2\langle\hat{\sigma}_{\alpha,1}^{22}\rangle- 1\big), \label{eq:B6} 
\end{align}
\begin{align}
&\frac d{dt}\langle\hat{\sigma}_{\alpha,1}^{22}\rangle =-\gamma_\alpha\langle\hat{\sigma}_{\alpha,1}^{22}\rangle+ig_\alpha\left(\langle\hat{a}^\dagger\hat{\sigma}_{\alpha,1}^{12}\rangle-\langle\hat{a}\hat{\sigma}_{\alpha,1}^{21}\rangle\right)\nonumber \\
&+i\Omega_\alpha\left(\langle\hat{\sigma}_{\alpha,1}^{12}\rangle-\langle\hat{\sigma}_{\alpha,1}^{21}\rangle\right). \label{eq:B2}
\end{align}
Here, we have introduced the complex frequencies $\tilde{\delta}_c = \delta_c+i\kappa/2$ and $\tilde{\delta}_\alpha = \delta_\alpha+i(\gamma_\alpha/2+\chi_\alpha)$.

We encounter many second-order mean-fields. The interactivity photon number $\langle\hat{a}^\dagger\hat{a}\rangle$ and the photon-photon correlation $\langle\hat{a}\hat{a}\rangle$ satisfy the equations  
\begin{align}
&\frac{d}{dt}\langle\hat{a}^\dagger\hat{a}\rangle = -\kappa\langle\hat{a}^\dagger\hat{a}\rangle + i\sum_{\alpha=1,2} N_\alpha g_\alpha\left(\langle\hat{a}\hat{\sigma}_{\alpha,1}^{21}\rangle -\langle\hat{a}^\dagger\hat{\sigma}_{\alpha,1}^{12}\rangle\right),\label{eq:B1}\\
&\frac{d}{dt}\langle\hat{a}\hat{a}\rangle = 2i\tilde{\delta}_c\langle\hat{a}\hat{a}\rangle - 2i\sum_{\alpha=1,2}N_\alpha g_\alpha\langle\hat{a}\hat{\sigma}_{\alpha,1}^{12}\rangle. \label{eq:B17}
\end{align}
The atom-photon correlations  $\langle\hat{a}\hat{\sigma}_{\alpha,1}^{12}\rangle$
, $\langle\hat{a}\hat{\sigma}_{\alpha,1}^{21}\rangle$, $\langle\hat{a}\hat{\sigma}_{\alpha,1}^{22}\rangle$ satisfy the equations
\begin{align}
&\frac d{dt}\langle\hat{a}\hat{\sigma}_{\alpha,1}^{12}\rangle=i(\tilde{\delta}_c+\tilde{\delta}_\alpha)\langle\hat{a}\hat{\sigma}_{\alpha,1}^{12}\rangle -ig_\alpha\langle\hat{a}\hat{a}\rangle \nonumber \\
& -i(N_\alpha-1)g_\alpha\langle\hat{\sigma}_{\alpha,1}^{12}\hat{\sigma}_{\alpha,2}^{12}\rangle +2ig_\alpha\langle\hat{a}\hat{a}\hat{\sigma}_{\alpha,1}^{22}\rangle  \nonumber \\
&-iN_{\alpha^{\prime}}g_{\alpha^{\prime}}\langle\hat{\sigma}_{\alpha,1}^{12}\hat{\sigma}_{\alpha^{\prime},1}^{12}\rangle+i\Omega_\alpha\big(2\langle\hat{a}\hat{\sigma}_{\alpha,1}^{22}\rangle-\langle\hat{a}\rangle\big), 
\end{align}
\begin{align}
&\frac d{dt}\langle\hat{a}\hat{\sigma}_{\alpha,1}^{21}\rangle = i(\tilde{\delta}_c-\tilde{\delta}^*_\alpha)\langle\hat{a}\hat{\sigma}_{\alpha,1}^{21}\rangle +ig_{\alpha}\big(\langle\hat{a}^{\dagger}\hat{a}\rangle \langle\hat{\sigma}_{\alpha,1}^{22}\rangle\big)\nonumber \\  & - -i(N_\alpha-1)g_\alpha\langle\hat{\sigma}_{\alpha,1}^{21}\hat{\sigma}_{\alpha,2}^{12} \rangle-2ig_\alpha\langle\hat{a}^\dagger\hat{a}\hat{\sigma}_{\alpha,1}^{22} \rangle\nonumber \\
& -iN_{\alpha^{\prime}}g_{\alpha^{\prime}}\langle\hat{\sigma}_{\alpha,1}^{21}\hat{\sigma}_{\alpha^{\prime},1}^{12}\rangle+i\Omega_{\alpha}\big(\langle\hat{a}\rangle-2\langle\hat{a}\hat{\sigma}_{\alpha,1}^{22}\rangle\big),\label{eq:B4}
\end{align}
\begin{align}
&\frac d{dt}\langle\hat{a}\hat{\sigma}_{\alpha,1}^{22}\rangle=(i\tilde{\delta}_c-\gamma_\alpha)\langle\hat{a}\hat{\sigma}_{\alpha,1}^{22}\rangle -i(N_\alpha-1)g_\alpha\langle\hat{\sigma}_{\alpha,1}^{22}\hat{\sigma}_{\alpha,2}^{12}\rangle \nonumber \\
&+ig_\alpha\big(\langle\hat{a}^\dagger\hat{a}\hat{\sigma}_{\alpha,1}^{12}\rangle-\langle\hat{a}\hat{a}\hat{\sigma}_{\alpha,1}^{21}\rangle\big)-iN_{\alpha^{\prime}}g_{\alpha^{\prime}}\langle\hat{\sigma}_{\alpha,1}^{22}\hat{\sigma}_{\alpha^{\prime},1}^{12}\rangle \nonumber \\
&+i\Omega_\alpha\big(\langle\hat{a}\hat{\sigma}_{\alpha,1}^{12}\rangle-\langle\hat{a}\hat{\sigma}_{\alpha,1}^{21}\rangle\big).
\label{eq:B8}
\end{align}
To simplify the notion, we retain the third-order mean-fields $\langle \hat{o}\hat{p}\hat{q}\rangle$, but approximate them with the following expression $\langle \hat{o}\hat{p}\hat{q}\rangle \approx\langle \hat{o}\rangle \langle \hat{p}\hat{q}\rangle +\langle \hat{p}\rangle \langle \hat{o}\hat{q}\rangle +\langle \hat{q}\rangle \langle \hat{o}\hat{p}\rangle -2\langle \hat{o}\rangle \langle \hat{p}\rangle \langle \hat{q}\rangle $ in the actual calculations.

The atom-atom correlations in the same sub-ensemble $\langle\hat{\sigma}_{\alpha,1}^{12}\hat{\sigma}_{\alpha,2}^{12}\rangle$ 
,$\langle\hat{\sigma}_{\alpha,1}^{21}\hat{\sigma}_{\alpha,2}^{12}\rangle$
,$\langle\hat{\sigma}_{\alpha,1}^{22}\hat{\sigma}_{\alpha,2}^{12}\rangle$
,$\langle\hat{\sigma}_{\alpha,1}^{22}\hat{\sigma}_{\alpha,2}^{22}\rangle$ satisfy the equations
\begin{align}
&\frac d{dt}\langle\hat{\sigma}_{\alpha,1}^{12}\hat{\sigma}_{\alpha,2}^{12}\rangle=2i\tilde{\delta}_\alpha\langle\hat{\sigma}_{\alpha,1}^{12}\hat{\sigma}_{\alpha,2}^{12}\rangle-2ig_\alpha\langle\hat{a}\hat{\sigma}_{\alpha,1}^{12}\rangle
\nonumber \\
&+4ig_\alpha\langle\hat{\sigma}_{\alpha,1}^{12}\hat{a}\hat{\sigma}_{\alpha,1}^{22}\rangle+2i\Omega_\alpha\big(2\langle\hat{\sigma}_{\alpha,1}^{22}\hat{\sigma}_{\alpha,2}^{12}\rangle-\langle\hat{\sigma}_{\alpha,1}^{12}\rangle\big),
\end{align}
\begin{align}
&\frac d{dt}\langle\hat{\sigma}_{\alpha,1}^{21}\hat{\sigma}_{\alpha,2}^{12}\rangle=- (\gamma_\alpha+2\chi_\alpha)\langle\hat{\sigma}_{\alpha,1}^{21}\hat{\sigma}_{\alpha,2}^{12}\rangle+ig_\alpha(\langle\hat{a}^\dagger\hat{\sigma}_{\alpha,1}^{12}\rangle \nonumber \\
&-\langle\hat{a}\hat{\sigma}_{\alpha,1}^{21}\rangle)+2ig_\alpha\big(\langle\hat{a}\hat{\sigma}_{\alpha,1}^{22}\hat{\sigma}_{\alpha,1}^{21}\rangle-\langle\hat{a}^\dagger\hat{\sigma}_{\alpha,2}^{12}\hat{\sigma}_{\alpha,1}^{22}\rangle\big)  \nonumber \\
& +i\Omega_\alpha\big(\langle\hat{\sigma}_{\alpha,1}^{12}\rangle-\langle\hat{\sigma}_{\alpha,1}^{21}\rangle\big) +2i\Omega_\alpha(\langle\hat{\sigma}_{\alpha,1}^{22}\hat{\sigma}_{\alpha,2}^{21}\rangle \nonumber \\
& -\langle\hat{\sigma}_{\alpha,1}^{22}\hat{\sigma}_{\alpha,2}^{12}\rangle),\label{eq:B10}
\end{align}
\begin{align}
&\frac{d}{dt}\langle\hat{\sigma}_{\alpha,1}^{22}\hat{\sigma}_{\alpha,2}^{12}\rangle=(i\tilde{\delta}_{\alpha}-\gamma_{\alpha})\langle\hat{\sigma}_{\alpha,1}^{22}\hat{\sigma}_{\alpha,2}^{12}\rangle-ig_{\alpha}\langle\hat{a}\hat{\sigma}_{\alpha,1}^{22}\rangle \nonumber \\
&+ig_\alpha\big(\langle\hat{a}^\dagger\hat{\sigma}_{\alpha,1}^{12}\hat{\sigma}_{\alpha,2}^{12}\rangle+2\langle\hat{a}\hat{\sigma}_{\alpha,1}^{22}\hat{\sigma}_{\alpha,2}^{22}\rangle-\langle\hat{\sigma}_{\alpha,1}^{12}\hat{a}\hat{\sigma}_{\alpha,1}^{21}\rangle\big)\nonumber \\
&+i\Omega_\alpha(2\langle\hat{\sigma}_{\alpha,1}^{22}\hat{\sigma}_{\alpha,2}^{22}\rangle+\langle\hat{\sigma}_{\alpha,1}^{12}\hat{\sigma}_{\alpha,2}^{12}\rangle-\langle\hat{\sigma}_{\alpha,1}^{21}\hat{\sigma}_{\alpha,2}^{12}\rangle\nonumber \\
& -\langle\hat{\sigma}_{\alpha,1}^{22}\rangle),
\end{align}
\begin{align}
&\frac d{dt}\langle\hat{\sigma}_{\alpha,1}^{22}\hat{\sigma}_{\alpha,2}^{22}\rangle=-2\gamma_\alpha\langle\hat{\sigma}_{\alpha,1}^{22}\hat{\sigma}_{\alpha,2}^{22}\rangle\nonumber \\
&+2ig_\alpha(\langle\hat{a}^\dagger\hat{\sigma}_{\alpha,1}^{22}\hat{\sigma}_{\alpha,2}^{12}\rangle-\langle\hat{\sigma}_{\alpha,1}^{21}\hat{\sigma}_{\alpha,1}^{22}\hat{a}\rangle) \nonumber \\
&+2i\Omega_\alpha(\langle\hat{\sigma}_{\alpha,1}^{22}\hat{\sigma}_{\alpha,2}^{12}\rangle-\langle\hat{\sigma}_{\alpha,1}^{22}\hat{\sigma}_{\alpha,2}^{21}\rangle).
\end{align}

The atom-atom correlations between different sub-ensembles $\langle\hat{\sigma}_{\alpha,1}^{12}\hat{\sigma}_{\alpha^{\prime},1}^{12}\rangle$,
$\langle\hat{\sigma}_{\alpha,1}^{21}\hat{\sigma}_{\alpha^{\prime},1}^{12} \rangle$,
$\langle\hat{\sigma}_{\alpha,1}^{22}\hat{\sigma}_{\alpha^{\prime},1}^{12}\rangle$,
$\langle\hat{\sigma}_{\alpha,1}^{21}\hat{\sigma}_{\alpha',1}^{22}\rangle$,
$\langle\hat{\sigma}_{\alpha,1}^{22}\hat{\sigma}_{\alpha^{\prime},1}^{22}\rangle$
satisfy the equations 
\begin{align}
&\frac d{dt}\langle\hat{\sigma}_{\alpha,1}^{12}\hat{\sigma}_{\alpha^{\prime},1}^{12}\rangle=i(\tilde{\delta}_\alpha+\tilde{\delta}_{\alpha'})\langle\hat{\sigma}_{\alpha,1}^{12}\hat{\sigma}_{\alpha^{\prime},1}^{12}\rangle-ig_\alpha\langle a\hat{\sigma}_{\alpha^{\prime},1}^{12}\rangle \nonumber \\
&-ig_{\alpha^{\prime}}\langle a\hat{\sigma}_{\alpha,1}^{12}\rangle +2ig_\alpha\langle\hat{\sigma}_{\alpha^{\prime},1}^{12}\hat{\sigma}_{\alpha,1}^{22}\hat{a}\rangle +2ig_{\alpha^{\prime}}\langle\hat{\sigma}_{\alpha^{\prime},1}^{22}\hat{a}\hat{\sigma}_{\alpha,1}^{12}\rangle \nonumber \\
&+i\Omega_\alpha\left(2\langle\hat{\sigma}_{\alpha,1}^{22}\hat{\sigma}_{\alpha^{\prime},1}^{12}\rangle-\langle\hat{\sigma}_{\alpha^{\prime},1}^{12}\rangle\right) +i\Omega_{\alpha^{\prime}}(2\langle\hat{\sigma}_{\alpha,1}^{12}\hat{\sigma}_{\alpha^{\prime},1}^{22}\rangle\nonumber \\
& -\langle\hat{\sigma}_{\alpha,1}^{12}\rangle),
\end{align}
\begin{align}
&\frac d{dt}\langle\hat{\sigma}_{\alpha,1}^{21}\hat{\sigma}_{\alpha^{\prime},1}^{12} \rangle=i(\tilde{\delta}_{\alpha'} - \tilde{\delta}_\alpha)\langle\hat{\sigma}_{\alpha,1}^{21}\hat{\sigma}_{\alpha^{\prime},1}^{12}\rangle+ig_{\alpha}\langle\hat{a}^{\dagger}\hat{\sigma}_{\alpha^{\prime},1}^{12}\rangle\nonumber \\
&-ig_{\alpha^{\prime}}\langle\hat{a}\hat{\sigma}_{\alpha,1}^{21}\rangle  -2ig_\alpha\langle\hat{\sigma}_{\alpha,1}^{22}\hat{a}^\dagger\hat{\sigma}_{\alpha^{\prime},1}^{12}\rangle+2ig_{\alpha^{\prime}}\langle\hat{\sigma}_{\alpha^{\prime},1}^{22}\hat{a}\hat{\sigma}_{\alpha,1}^{21}\rangle\nonumber \\
& +i\Omega_\alpha\big(\langle\hat{\sigma}_{\alpha',1}^{12}\rangle-2\langle\hat{\sigma}_{\alpha,1}^{22}\hat{\sigma}_{\alpha',1}^{12}\rangle\big)+i\Omega_{\alpha'}(2\langle\hat{\sigma}_{\alpha,1}^{21}\hat{\sigma}_{\alpha',1}^{22}\rangle\nonumber \\
& -\langle\hat{\sigma}_{\alpha,1}^{21}\rangle),
\end{align}

\begin{align}
&\frac d{dt}\langle\hat{\sigma}_{\alpha,1}^{22}\hat{\sigma}_{\alpha^{\prime},1}^{12}\rangle=(i\tilde{\delta}_{\alpha'}-\gamma_\alpha)\langle\hat{\sigma}_{\alpha,1}^{22}\hat{\sigma}_{\alpha^{\prime},1}^{12}\rangle-ig_{\alpha^{\prime}}\langle\hat{a}\hat{\sigma}_{\alpha,1}^{22}\rangle\nonumber \\
&+ig_\alpha\big(2\langle\hat{\sigma}_{\alpha^{\prime},1}^{22}\hat{a}\hat{\sigma}_{\alpha,1}^{22}\rangle+\langle\hat{\sigma}_{\alpha,1}^{12}\hat{a}^{\dagger}\hat{\sigma}_{\alpha^{\prime},1}^{12}\rangle-\langle\hat{\sigma}_{\alpha^{\prime},1}^{12}\hat{\sigma}_{\alpha,1}^{21}\hat{a}\rangle) \nonumber \\
&+i\Omega_\alpha\big((\hat{\sigma}_{\alpha,1}^{12}\hat{\sigma}_{\alpha^{\prime},1}^{12})-(\hat{\sigma}_{\alpha,1}^{21}\hat{\sigma}_{\alpha^{\prime},1}^{12})\big)+i\Omega_{\alpha^{\prime}}(2\langle\hat{\sigma}_{\alpha,1}^{22}\hat{\sigma}_{\alpha^{\prime},1}^{22}\rangle\nonumber \\
& -\langle\hat{\sigma}_{\alpha,1}^{22}\rangle),
\end{align}
\begin{align}
&\frac{d}{dt}\langle\hat{\sigma}_{\alpha,1}^{21}\hat{\sigma}_{\alpha',1}^{22}\rangle=-(i\tilde{\delta}^*_{\alpha}+\gamma_{\alpha'})\langle\hat{\sigma}_{\alpha,1}^{21}\hat{\sigma}_{\alpha',1}^{22}\rangle +ig_{\alpha}(\langle\hat{a}^{\dagger}\hat{\sigma}_{\alpha^{\prime},1}^{22}\rangle\nonumber \\
&-2\langle\hat{a}^{\dagger}\hat{\sigma}_{\alpha^{\prime},1}^{22}\hat{\sigma}_{\alpha,1}^{22}\rangle)+ig_{\alpha^{\prime}}\Big(\langle\hat{\sigma}_{\alpha,1}^{21}\hat{a}^{\dagger}\hat{\sigma}_{\alpha^{\prime},1}^{12}\rangle-\langle\hat{\sigma}_{\alpha^{\prime},1}^{21}\hat{a}\hat{\sigma}_{\alpha,1}^{21}\rangle\Big) \nonumber \\
&+i\Omega_\alpha\Big(\langle\hat{\sigma}_{\alpha^{\prime},1}^{22}\rangle-2\langle\hat{\sigma}_{\alpha,1}^{22}\hat{\sigma}_{\alpha^{\prime},1}^{22}\rangle\Big)+i\Omega_{\alpha^{\prime}}(\langle\hat{\sigma}_{\alpha,1}^{21}\hat{\sigma}_{\alpha^{\prime},1}^{12}\rangle\nonumber \\
& -\langle\hat{\sigma}_{\alpha,1}^{21}\hat{\sigma}_{\alpha^{\prime},1}^{21}\rangle),
\end{align}
\begin{align}
&\frac{d}{dt}\langle\hat{\sigma}_{\alpha,1}^{22}\hat{\sigma}_{\alpha^{\prime},1}^{22}\rangle=- (\gamma_{\alpha}+\gamma_{\alpha^{\prime}})\langle\hat{\sigma}_{\alpha,1}^{22}\hat{\sigma}_{\alpha^{\prime},1}^{22}\rangle +ig_{\alpha}(\langle\hat{a}^{\dagger}\hat{\sigma}_{\alpha^{\prime},1}^{22}\hat{\sigma}_{\alpha,1}^{12}\rangle\nonumber \\
&-\langle\hat{\sigma}_{\alpha^{\prime},1}^{22}\hat{a}\hat{\sigma}_{\alpha,1}^{21}\rangle)+ig_{\alpha^{\prime}}\big(\langle\hat{\sigma}_{\alpha,1}^{22}\hat{a}^{\dagger}\hat{\sigma}_{\alpha^{\prime},1}^{12}\rangle-\langle\hat{\sigma}_{\alpha^{\prime},1}^{21}\hat{a}\hat{\sigma}_{\alpha,1}^{22}\rangle\big)\nonumber \\
&+i\Omega_{\alpha}\big(\langle\hat{\sigma}_{\alpha,1}^{12}\hat{\sigma}_{\alpha^{\prime},1}^{22}\rangle-\langle\hat{\sigma}_{\alpha,1}^{21}\hat{\sigma}_{\alpha^{\prime},1}^{22}\rangle\big) +i\Omega_{\alpha^{\prime}}(\langle\hat{\sigma}_{\alpha,1}^{22}\hat{\sigma}_{\alpha^{\prime},1}^{12}\rangle\nonumber \\
& -\langle\hat{\sigma}_{\alpha,1}^{22}\hat{\sigma}_{\alpha^{\prime},1}^{21}\rangle).
\end{align}

We have also encountered the mean-fields  $\langle\hat{\sigma}_{\alpha^{\prime},1}^{12}\rangle$,
$\langle\hat{\sigma}_{\alpha^{\prime},1}^{22}\rangle$,
$\langle\hat{a}^{\dagger}\hat{\sigma}_{\alpha^{\prime},1}^{12}\rangle$,
$\langle\hat{a}^{\dagger}\hat{\sigma}_{\alpha^{\prime},1}^{21}\rangle$,
$\langle\hat{a}^\dagger\hat{\sigma}_{\alpha^{\prime},1}^{22}\rangle$,
$\langle\hat{\sigma}_{\alpha^{\prime},1}^{21}\hat{\sigma}_{\alpha^{\prime},2}^{12}\rangle$,
$\langle\hat{\sigma}_{\alpha^{\prime},1}^{22}\hat{\sigma}_{\alpha^{\prime},2}^{21}\rangle$,
$\langle\hat{\sigma}_{\alpha^{\prime},1}^{22}\hat{\sigma}_{\alpha^{\prime},2}^{22}\rangle$. Since they are the complex conjugation of the mean-fields considered above, we do not present the equations for these terms. 

\section{\label{sec:Extra}Extra Numerical Results}

In this Appendix, we show the extra numerical results to complement those given in the main text.

\begin{figure}[htbp]
\begin{centering}
\includegraphics[width=0.5\textwidth]{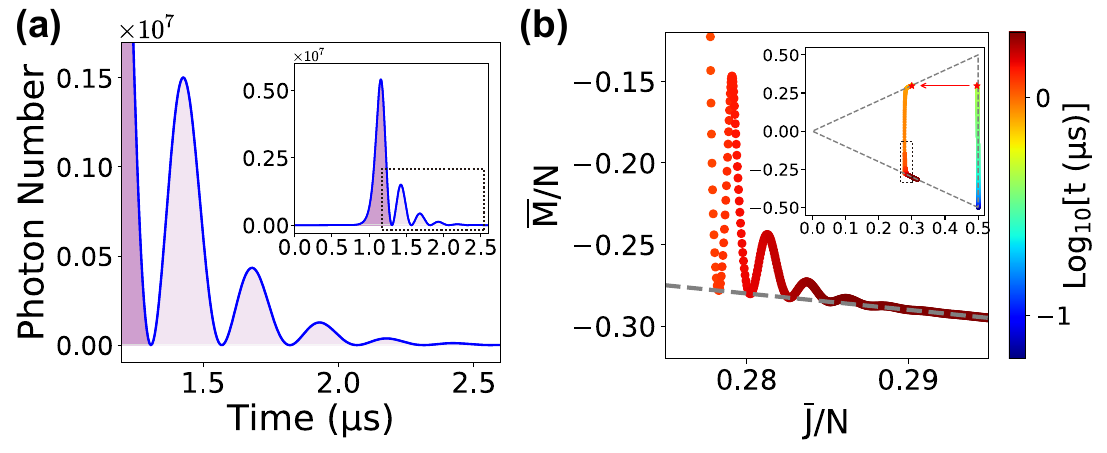} 
\par\end{centering}
\caption{\label{fig:rabi} Rabi oscillations dynamics for the system with $2\times 10^7$ atoms. Panel (a) shows the dynamics of the intra-cavity photon number [inset, the recapitulation of the upper part of Fig.~\ref{fig:pulses}(e)], and the zoom-in of the regions after the dominated peak. Panel (b) shows the dynamics of the atomic ensemble in the Dicke states space in the staggered picture [inset, the recapitulation of Fig.~\ref{fig:pulses}(a)], and the zoom-in of the region after the vertical decay.}
\end{figure}

\subsection{Rabi Oscillations Dynamics}

In Fig.~\ref{fig:pulses}, we have examined the superradiant pulses from the system with $2\times 10^7$ atoms, and identified several small pulses after the dominated one  for longer driving pulse, see the upper part of Fig.~\ref{fig:pulses}(e) and the inset of Fig.~\ref{fig:rabi}(a). To understand these small pulses, we examine carefully the dynamics of the atomic ensemble in the Dicke states space after the vertical decay [Fig.~\ref{fig:rabi}(b)]. We find that the atomic ensemble is firstly excited to the Dicke states with larger $\bar{M}$ by re-absorbing the photons and then decay from these states to the lower boundary by emitting photons, and the same dynamics repeats with however reduced maximal excitation and photons. Thus, the result shows here can be attributed to the Rabi oscillations as often observed for the system in the crossover or strong coupling regime~\citep{Norcia2016}.

\begin{figure}[htbp]
\begin{centering}
\includegraphics[width=0.5\textwidth]{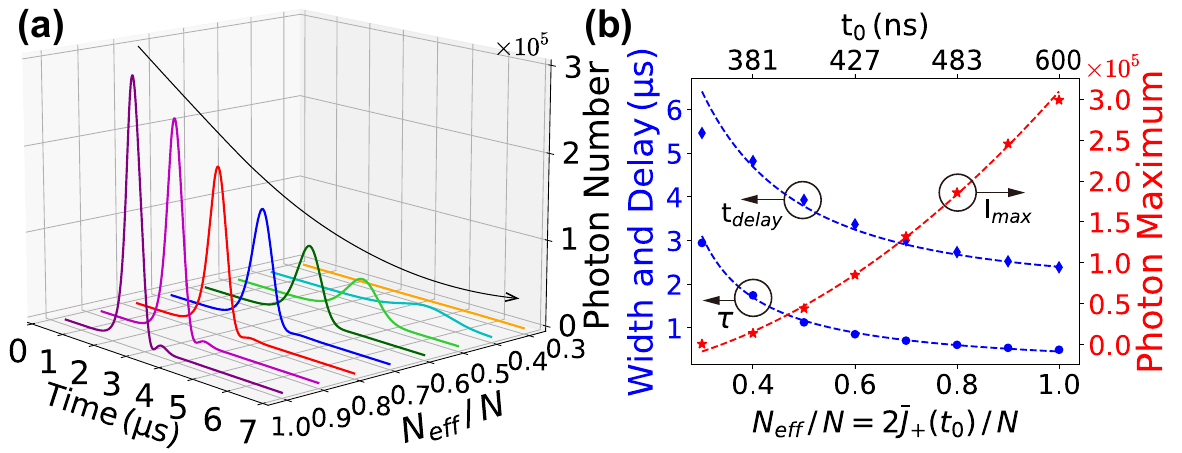} 
\par\end{centering}
\caption{\label{fig:pulses-few-atoms} Delayed superradiant pulses for the system with  $N=2\times 10^{6}$ atoms.   Panel (a) shows the pulse as function of the ratio $r=N_{eff}/N$ of the effective number of atoms $N_{eff}=2\bar{J}_+(t_0)$  and the total number of atoms $N$.  
Panel (b) shows the  maximum $I_{max}$ , delayed time $t_{delay}$, width $\tau$  of the pulses as function of the ratio $r$, which are fitted with expressions $\sim r^2,\sim {\rm ln}r/r, \sim 1/r $, respectively. Here, $N_{eff}$ can be simply varied by the length of the driving pulses. }
\end{figure}

\subsection{Delayed Superradiant Pulses for System in the Weak Coupling Regime}

In Fig.~\ref{fig:pulses} (f) of the main text, we have studied the characteristics of the dominated superradiant pulses for a system with $N=2\times 10^7$ atoms, and found that the pulse maximum scales linearly with the ratio $r=N_{eff}/N$ of the effective number of atoms $N_{eff}$  and the total number of atoms $N$.  We attribute the observed linear scaling to the systems in the strong coupling regime~\citep{Gogyan2020}, and argue that the quadratic scaling should be observed for the systems in the weak coupling regime. To verify our argument, we study  the delayed superradiance for the  system with fewer number of atoms ($N=2\times 10^6$ )[ Fig. ~\ref{fig:pulses-few-atoms} (a)], and examined the dependence of the maximum, delayed time, width of the pulses on the ratio  $r$ [ Fig. ~\ref{fig:pulses-few-atoms} (b)]. Indeed, we find only single pulses in most case, and a quadratic scaling of the pulse maximum with the ratio $r$  or the effective number of atoms $N_{eff}$.

\begin{figure}[htbp]
\begin{centering}
\includegraphics[width=0.5\textwidth]{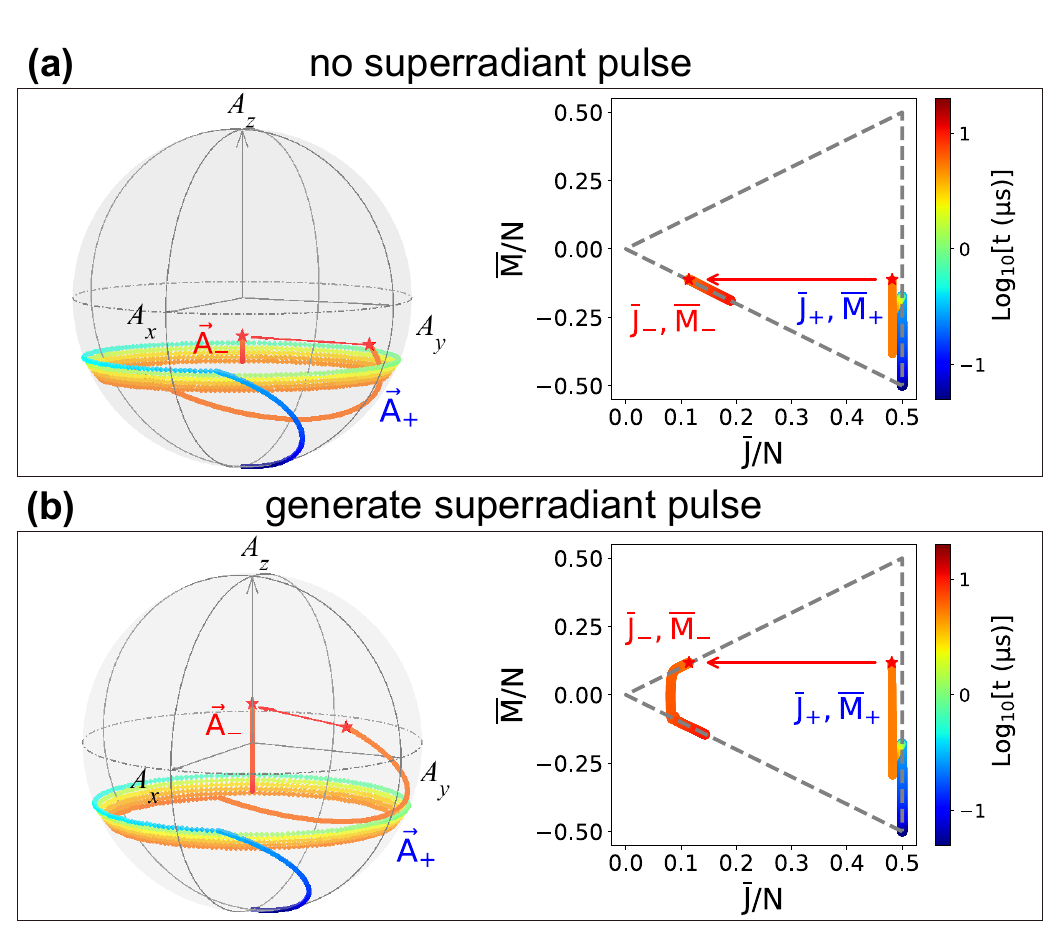} 
\par\end{centering}
\caption{\label{fig:Pulse-large}Results similar to Fig.~\ref{fig:Ramsey} (b,c) but for the system with  large frequency detuning $\delta=2\pi\times 1150$ kHz , and  $\delta=2\pi\times1155$ kHz , which does not and does generate a delayed superradiant pulse, respectively .}
\end{figure}

\subsection{Atomic Ensemble Dynamics During Ramsey Measurement for Large Frequency Detuning } 

In the section~\ref{sec:ramsey spectroscopy} of the main text, we show that the accumulated phases for the the peaks and zeros in the superradiant Ramsey spectrum are not exact multiples of $\pi$, and the signal amplitude decreases for larger frequency detuning. To understand the reasons behind, we show the dynamics of the atomic ensemble for two large frequency detuning in Fig.~\ref{fig:Pulse-large}, where the superradiant pulse is present and absent correspondingly. The numerical results show that the axis of the Bloch vector rotation becomes tilted from the x-axis, the rotation becomes faster but less strong, and the Bloch vector circles more times during the free procession. At the same time, in the Dicke states space, the atomic ensemble climb to the Dicke states with much lower excitation quantum number $\bar{M}$ due to the laser excitation. Thus, the relationship between the final state after the second laser pulse and the generated superradiant pulses is different from that in the resonant condition, which leads to the observed phenomena.

\end{document}